\documentclass[fleqn,usenatbib]{mnras}

\usepackage{newtxtext,newtxmath}

\usepackage[T1]{fontenc}

\DeclareRobustCommand{\VAN}[3]{#2}
\let\VANthebibliography\thebibliography
\def\thebibliography{\DeclareRobustCommand{\VAN}[3]{##3}\VANthebibliography}

\usepackage{graphicx}	
\usepackage{amsmath}	
\usepackage{amssymb}	

\title[Relative depolarization of the black hole photon ring in GRMHD]{Relative depolarization of the black hole photon ring in GRMHD models of Sgr A* and M87*}

\author[Jim\'{e}nez-Rosales, Dexter, Ressler, Tchekhovskoy, et al.]{
A. Jim\'{e}nez-Rosales,$^{1,2}$\thanks{a.jimenez@science.ru.nl}
J. Dexter,$^{3,2}$\thanks{jason.dexter@colorado.edu}
S. M. Ressler,$^{4}$
A. Tchekhovskoy,$^{5}$
M. Baub\"{o}ck,$^{2}$
\newauthor
Y. Dallilar,$^{2}$ P. T. de Zeeuw,$^{2,6}$  A. Drescher,$^{2,7}$ F. Eisenhauer,$^{2}$ S. von Fellenberg,$^{2}$ F. Gao,$^{2}$ \newauthor
R. Genzel,$^{2,8}$ S. Gillessen,$^{2}$ M. Habibi,$^{2}$ T. Ott,$^{2}$ J. Stadler,$^{2}$ O. Straub,$^{2}$ F. Widmann$^{2}$
\\
$^{1}$Department of Astrophysics, IMAPP, Radboud University, 6500 GL Nijmegen, The Netherlands\\
$^{2}$Max Planck Institute for Extraterrestrial Physics (MPE), Giessenbachstr. 1, 85748 Garching, Germany\\
$^{3}$JILA and Department of Astrophysical and Planetary Sciences, University of Colorado, Boulder, CO 80309, USA\\
$^{4}$Kavli Institute for Theoretical Physics, University of California Santa Barbara, Kohn Hall, Santa Barbara, CA 93107, USA\\
$^{5}$Center for Interdisciplinary Exploration \& Research in Astrophysics (CIERA), Physics \& Astronomy, Northwestern University,\\ Evanston, IL 60202, USA\\
$^{6}$Sterrewacht Leiden, Leiden University, Postbus 9513, 2300 RA Leiden, The Netherlands\\
$^{7}$Department of Physics, Technical University Munich, James-Franck-Straße 1, 85748 Garching, Germany\\
$^{8}$Departments of Physics and Astronomy, Le Conte Hall, University of California, Berkeley, CA 94720, USA
}

\date{Accepted XXX. Received YYY; in original form ZZZ}

\pubyear{2020}

\begin{document}
\label{firstpage}
\pagerange{\pageref{firstpage}--\pageref{lastpage}}
\maketitle

\begin{abstract}
Using general relativistic magnetohydrodynamic simulations of accreting black holes, we show that a suitable subtraction of the linear polarization per pixel from total intensity images can enhance the photon ring features. 
We find that the photon ring is typically a factor of $\simeq 2$ less polarized than the rest of the image. This is due to a combination of plasma and general relativistic effects, as well as magnetic turbulence.
When there are no other persistently depolarized image features, adding the subtracted residuals over time results in a sharp image of the photon ring.
We show that the method works well for sample, viable GRMHD models of Sgr A* and M87*, where measurements of the photon ring properties would provide new measurements of black hole mass and spin, and potentially allow for tests of the ``no-hair'' theorem of general relativity.
\end{abstract}




\section{Introduction}

Long baseline interferometry techniques have the power to resolve and explore the innermost regions of accretion flows around supermassive black holes. Interferometric measurements made with GRAVITY at the Very Large Telescope \citep{eisenhauer2008,paumard2008,gravity2017} and the Event Horizon Telescope \citep[EHT; ][]{doeleman2008,doeleman2012,fish2011}, have offered unprecedented capability to study these systems at angular scales of tens of microarcseconds ($\mu$as). That resolution is comparable to the size of the event horizon on the sky for the Galactic center black hole, Sagittarius A* (Sgr A*), and the supermassive black hole in M87, M87*. The observations probe not only the geometrical and physical properties of matter around the compact object, but also allow for tests in the strong-field regime of general relativity \citep[][]{gravity2018,gravity2018flare,gravity2020_michi,gravity2020_precesion,eht2019i,eht2019ii,eht2019iii,eht2019iv,eht2019v,eht2019vi}.

Theoretical studies of low-luminosity accretion onto black holes encompass analytic models \citep[e.g., ][]{ichimaru1977,rees1982,narayan1994,yuan2003}, semi analytic calculations \citep[e.g., ][]{falcke2000,bromley2001,broderick2005,broderick2006} and numerical simulations. The latter include general relativistic magnetohydrodynamic (GRMHD)  calculations, where the equations of ideal magnetohydrodynamics are written, solved and self-consistently evolved  
taking into account general relativistic effects \citep[e.g., ][]{gammie2003,devilliers2003, noble2006,tchekhovskoy2011,narayan2012,shiokawa2013}.

While predicted total intensity images of emission arising at event horizon scales are useful in studying the properties of the accretion flow \citep[e.g. size, shape, variability; ][]{dexter2009mmvar,moscibrodzka2009,chan2015,anantua2020,dexter2020}, they are often dominated by light bending and Doppler boosting.
Polarization information has proven to be a key element in complementing these studies, setting limits on model quantities such as accretion rate and electron temperature from observables like linear polarization fraction \citep[e.g.,][]{aitken2000,agol2000,quataert2000,bower2017}, rotation measure \citep{bower2003,marrone2006,marrone2007,kuo2014,bower2018} and spatially resolved magnetic field structure \citep[e.g., ][]{johnson2015,moscibrodzka2017,jimenez-rosales2018,gravity2020_pol}.

The emission around a black hole is gravitationally lensed. At event horizon scales, the strength of this effect is such that it produces a sequence of strongly lensed images of the surrounding emission \citep[e.g.,][]{cunningham1973,luminet1979,viergutz1993} that result in an annulus of enhanced brightness (we henceforth refer to all indirect images as the ``photon ring'').  
Analytic studies show that if the emission around the black hole is optically thin, the photon ring may show universal features both in total intensity and polarization that are completely governed by general relativity \citep{johnson2019,himwich2020}.

The EHT Collaboration's published images of M87* show a bright ring of emission surrounding a dark interior \citep{eht2019iv}. The presence of a photon ring is expected within this emission, surrounding the black hole shadow \citep{falcke2000}. Measuring its properties, including the size and shape, provide new measurements of the mass and spin of the black hole and could even test the ``no-hair'' theorem of general relativity \citep[e.g.,][]{johannsen2010}. Typically, the photon ring contributes $\simeq 5-10\%$ of the total flux in GRMHD models \citep[e.g.,][]{ricarte2015}. The challenge is to separate this feature from surrounding direct emission. \citet{psaltis2015} showed that the edge of the photon ring feature is generally sharp, which may help with its extraction. This may also be possible using extremely long baselines (e.g., with one telescope in space), if stochastic source features can be smoothed by coherently averaging \citep{johnson2019}.

Here, we use GRMHD calculations to show that the polarized radiation may provide another method. In many current GRMHD models, the photon ring is enhanced when suitably subtracting the linear polarized flux per pixel from the total intensity (Section \ref{sec:method}). 
We find that the photon ring is persistently less polarized relative to rest of the image by a factor of $\simeq 2-2.5$ (Section \ref{sec:num_properties}). The relative depolarization is due to a combination of plasma and general relativistic effects, as well as magnetic field disorder. We discuss our results and conclude in Section \ref{sec:summary}.


\section{Photon ring extraction}
\label{sec:method}

\begin{figure*}
\centering
\includegraphics[trim = 0cm 0cm 0cm 0cm, clip=true,width=1\textwidth]{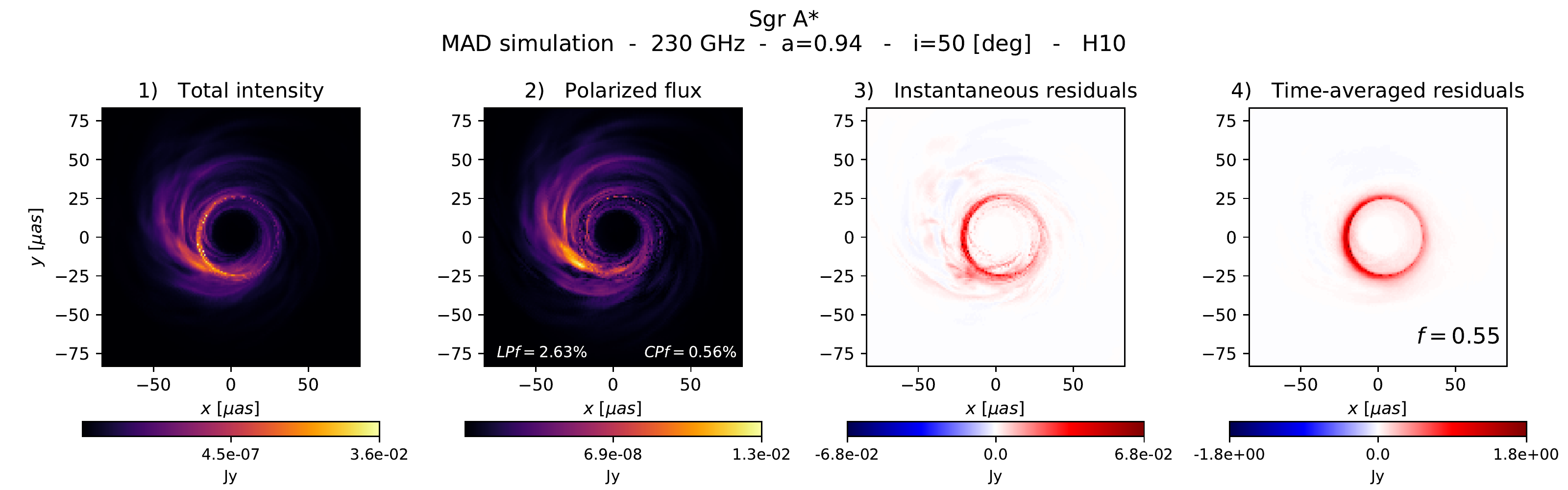}
\includegraphics[trim = 0cm 0cm 0cm 0cm, clip=true,width=1\textwidth]{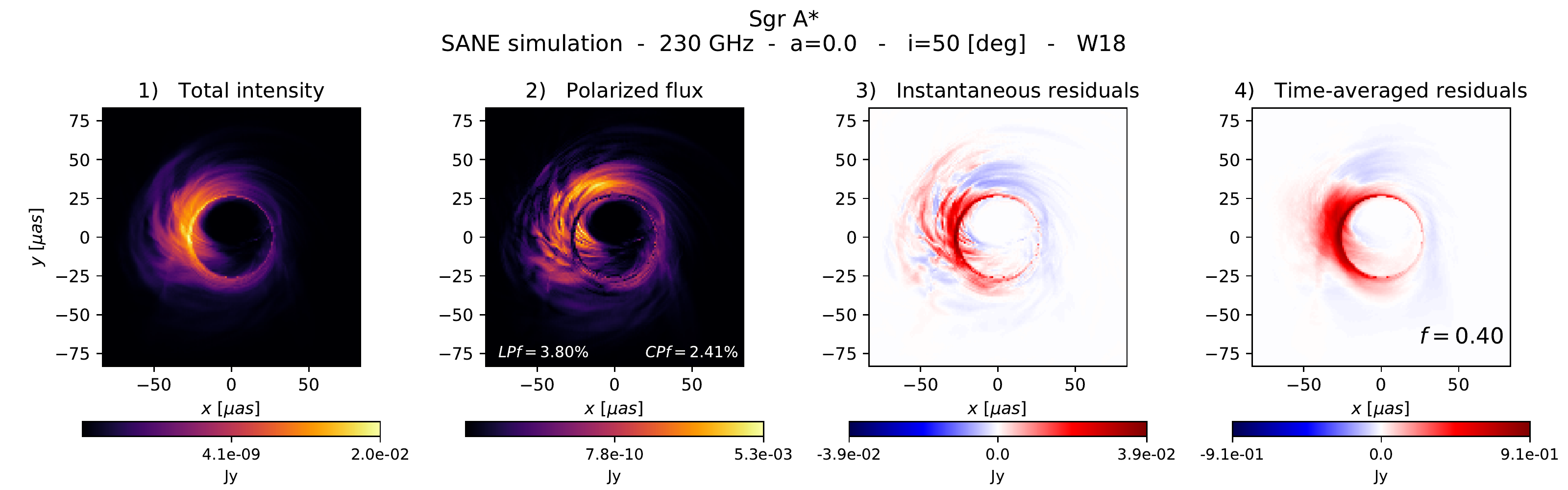}
 \caption{ Total intensity, polarized flux and residual images of Sgr A* at 230 GHz. Top row: MAD simulation with spin $a=0.94$, inclination $i=50^{\circ}$ and turbulent electron heating (H10). Bottom row: SANE simulation with spin $a=0.0$, inclination $i=50^{\circ}$ and magnetic reconnection as the electron heating mechanism (W18). Panels 1 and 2 show respectively the total intensity $I$ and the polarized flux $LP$ for a snapshot of the simulations. Emission is shown in false color. Higher emission zones are denoted by lighter colors. Both images exhibit similar morphologies, with the difference that the polarized flux image has many more dark areas in the accretion flow due to depolarization. The photon ring is a prominent feature in both and appears less polarized compared to the rest of the image. The net fractional LP and CP in the image are indicated in Panel 2. 3) Instantaneous residuals $I-1/f\ LP$ for a snapshot. 
 4) Time-averaged residuals over thirty frames. The snapshots are separated by $10 \ GM/c^3$. If the depolarized image features change over time except at the photon ring, the residuals average out everywhere else, enhancing the ring in the residual image. The factor $f$ indicated at the bottom of panel 4 is such that the residuals in the image are minimized.
 }
     \label{fig:snapshots_GRMHD}
\end{figure*}

In this work we use two long duration, 3D GRMHD simulations of black hole accretion flows described in \citet{dexter2020}: one in a strongly magnetized (magnetically arrested disc, MAD) regime and the other in a weakly magnetized (standard and normal evolution, SANE) one. The simulation's framework allows for the independent, self-consistent evolution of four electron internal energy densities in pair with that of the single MHD fluid \citep{ressler2015}. Here, we use a turbulent heating prescription based on gyrokinetic theory \citep[H10,][]{howes2010} for the MAD model and electron heating in magnetic reconnection events from particle-in-cell simulations \citep[W18,][]{werner2018} for the SANE model.

Both simulations are run with the \textsc{harmpi}\footnote{\url{https://github.com/atchekho/harmpi}} code \citep{tchekhovskoy2019}. 
The initial condition is that of a Fishbone-Moncrief torus \citep{fishbone1976} with inner radius at $r_{\rm in}=12\ r_g$ ($r_g=GM/c^2$), pressure maximum radius $r_{\rm max}=25\ r_g$, threaded with a single poloidal loop of magnetic field and black hole spin parameter of $a=0$ and $a=0.9375$ for the SANE and MAD cases respectively. 

In the MAD case, the simulation was run for a time of $6\times10^4 \ r_g/c$, each snapshot spaced by $10 \, r_g/c$, establishing inflow equilibrium out to $r \simeq 90\ r_g$.
In the SANE case, the simulation is studied at late times ($1.7\times10^4 \,r_g/c$, each snapshot spaced by $10 \, r_g/c$ as well), once radii $r\gtrsim 100\ r_g$ have reached inflow equilibrium. By doing this, we avoid ``artificial'' depolarization in the images from external Faraday rotation from zones far out in the accretion flow that are not yet in equilibrium. 
See \citet{dexter2020} for more details.

We calculate post-processed images using the general relativistic ray-tracing public code \textsc{grtrans}\footnote{\url{https://github.com/jadexter/grtrans}} \citep{dexter2009,dexter2016}. 
The electron temperature is taken directly from the GRMHD electron internal energy density, $k T_e = (\gamma_e - 1) m_p u_e / \rho$, where we use $\gamma_e = 4/3$ and have assumed a composition of pure ionized hydrogen. The mass of the black hole is set to $4\times 10^6\ \rm{M}_\odot$, where $\rm{M}_\odot$ indicates the solar mass unit, and the mass accretion rate is scaled until the observed flux density at $230$ GHz matches that of Sgr A* at this frequency \citep[$\simeq 3$ Jy, e.g.,][]{dexter2014,bower2015}. In our calculations, we exclude emission from regions where $\sigma = u_B/\rho c^2 > 1$, with $u_B = b^2/4\pi$ the magnetic energy density. We calculate observables at $230$ GHz and a viewing orientation of $i=50$ degrees. 
The image resolution is $192$ pixels over a $42 \ r_g$ ($\sim 210 \, \mu$as) field of view (a discussion of resolution effects can be found in Appendix \ref{appendix:resolution_effects}).

Panels 1 and 2 in the top and bottom rows of Fig.~\ref{fig:snapshots_GRMHD} show total intensity (Stokes $I$) and linear polarized flux, $LP=\sqrt{Q^2+U^2}$, of snapshot images of the MAD and SANE simulations respectively, where the Stokes parameters $Q$ and $U$ denote the linear polarization states at $0/90$ and $\pm45$ degrees at each pixel. 
In both cases, the total intensity and linear polarized flux images exhibit similar morphologies. 
We note as well that the images present a sharp, bright photon ring with many depolarized areas on it.
The net linear and circular polarization fractions (LPf\footnote{ LPf=$\sqrt{\langle \rm{Q}\rangle^2+\langle \rm{U}\rangle^2}/\langle \rm{I}\rangle$, where $\langle\rangle$ denotes the sum over all pixels.} and CPf\footnote{ CPf=$\langle \rm{V}\rangle/\langle \rm{I}\rangle$, with $V$ the circular polarization Stokes parameter.} respectively) in the image are indicated in Panel 2.

Suitably subtracting the polarized flux from the total intensity per pixel results in a residual image with a clearer photon ring. 
We introduce a scaling factor $f$ that accounts for the difference between the image's polarized flux and total intensity\footnote{ Since light might not be completely linearly polarized, $I\geq \rm{LP}$, so that $\langle I\rangle \geq \langle \rm{LP}\rangle$. To account for this difference, the factor $f$ should then be similar to $\langle\rm{LP}\rangle / \langle I\rangle$.}. 
The residual image is then $I - 1/f \ LP$. An example of the ``instantaneous'' (one snapshot) residuals for the values shown in Panels 1 and 2 is shown in Panel 3 in the top and bottom rows of Fig.~\ref{fig:snapshots_GRMHD}. 
Here we have run through a set of values of $f$ ($0.3-1.0$ spaced every $0.05$ and $f=\langle\rm{LP}\rangle / \langle I\rangle$) and chosen the value which minimizes the sum of the residuals outside of a thin ring on the sky 
with a width from $4.8-5.5 \, r_g$. This is close to expected position of the photon ring ($\sqrt{27} \, r_g$ for a Schwarzschild black hole and $\simeq 5 \, r_g \pm 4\%$ and nearly circular across all spins, \citealt{johannsen2010}) and where most of the emission is contained. 
In our calculations we consider as ``photon ring pixels'' those that lie within this ring. 
The images are inspected to ensure that all of the photon ring pixels are included. Save for small translations of the center of the ring due to spin changes between the simulations, the pixel selection radii remains the same. 
Note that our definition of photon ring pixels includes foreground ``contamination'' from the direct image, but this is generally a minor contribution. 

The total intensity and LP images vary in time due to turbulent fluctuations in the accretion flow. These variations are strongly correlated, except that the LP image varies somewhat more. This is because the direction of the polarization vector depends strongly on the underlying magnetic turbulence and/or Faraday effects, resulting in transient depolarization of some image pixels.
The photon ring, however, remains visible and less polarized over time.
The stability of the photon ring properties implies that the extraction technique benefits from source variability. Adding residual images tends to average out the residuals everywhere but the photon ring, increasing its contrast.
Panel 4 in the top and bottom rows of Fig.~\ref{fig:snapshots_GRMHD} shows the cumulative residual images over thirty frames of our simulations for each case, starting from $5.1\times10^3\,r_g/c$ for the MAD simulation and $1.7\times10^4\,r_g/c$ for the SANE simulation. 
This corresponds to a time span of $\sim1.7$ hr for Sgr A* and $\sim100$ days for M87*. The best $f$ value that minimizes the residuals in the time-averaged image is indicated in panel 4. 
It can be seen that an optimal choice of $f$ leads to the residuals successfully averaging out over time, sharpening the photon ring.


\section{Numerical properties of the photon ring}
\label{sec:num_properties}

\begin{figure*}
\centering
\includegraphics[trim = 3cm 8.8cm 3cm 0cm, clip=true,width=1\textwidth]{./figures_final/sgra_mad_gmin1_all_effects_230i50a94_panels.pdf}
\includegraphics[trim = 0cm 0cm 0cm 1.55cm, clip=true, width=0.45\textwidth]{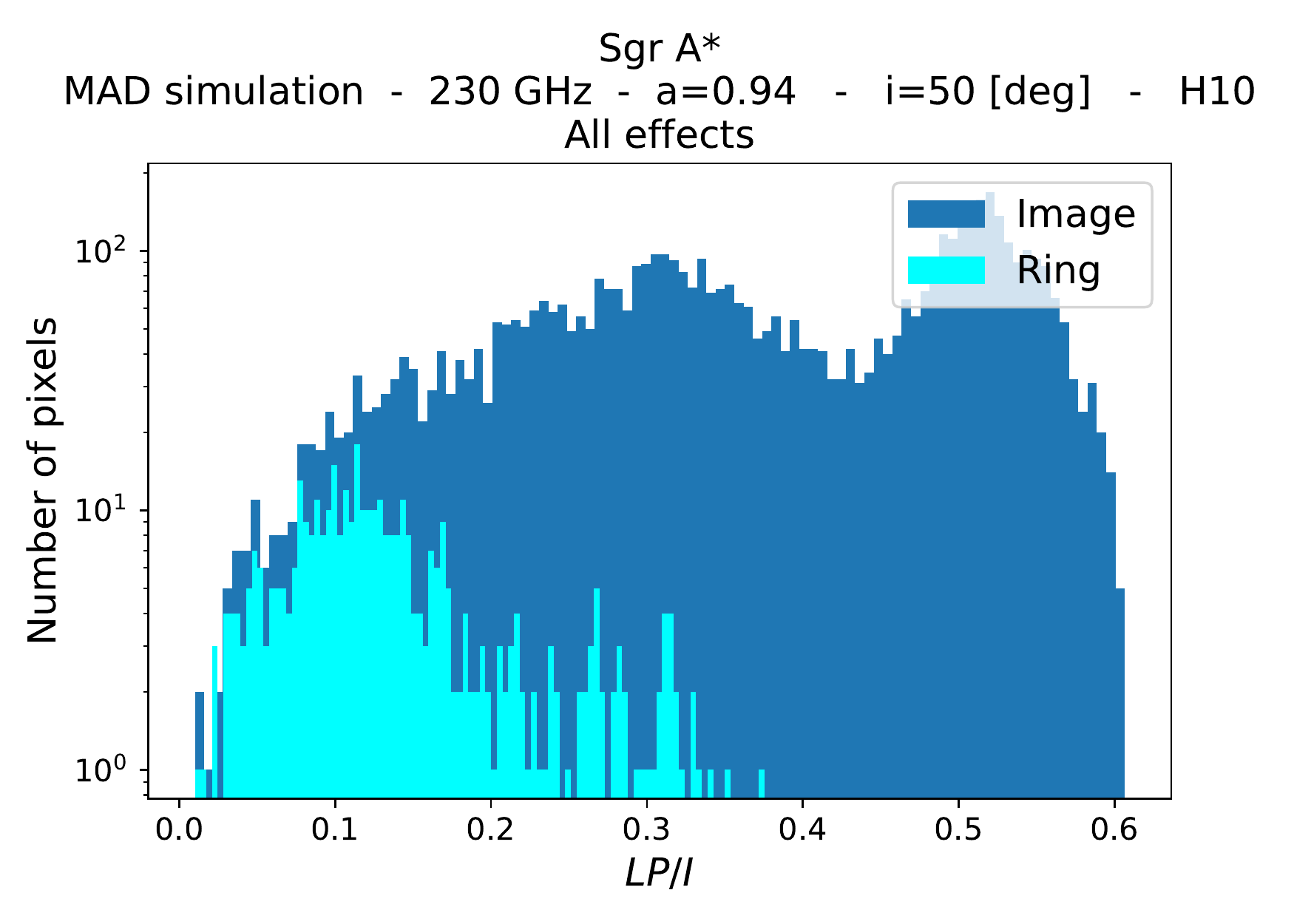}
\includegraphics[trim = 0cm 0cm 0cm 1.55cm, clip=true, width=0.45\textwidth]{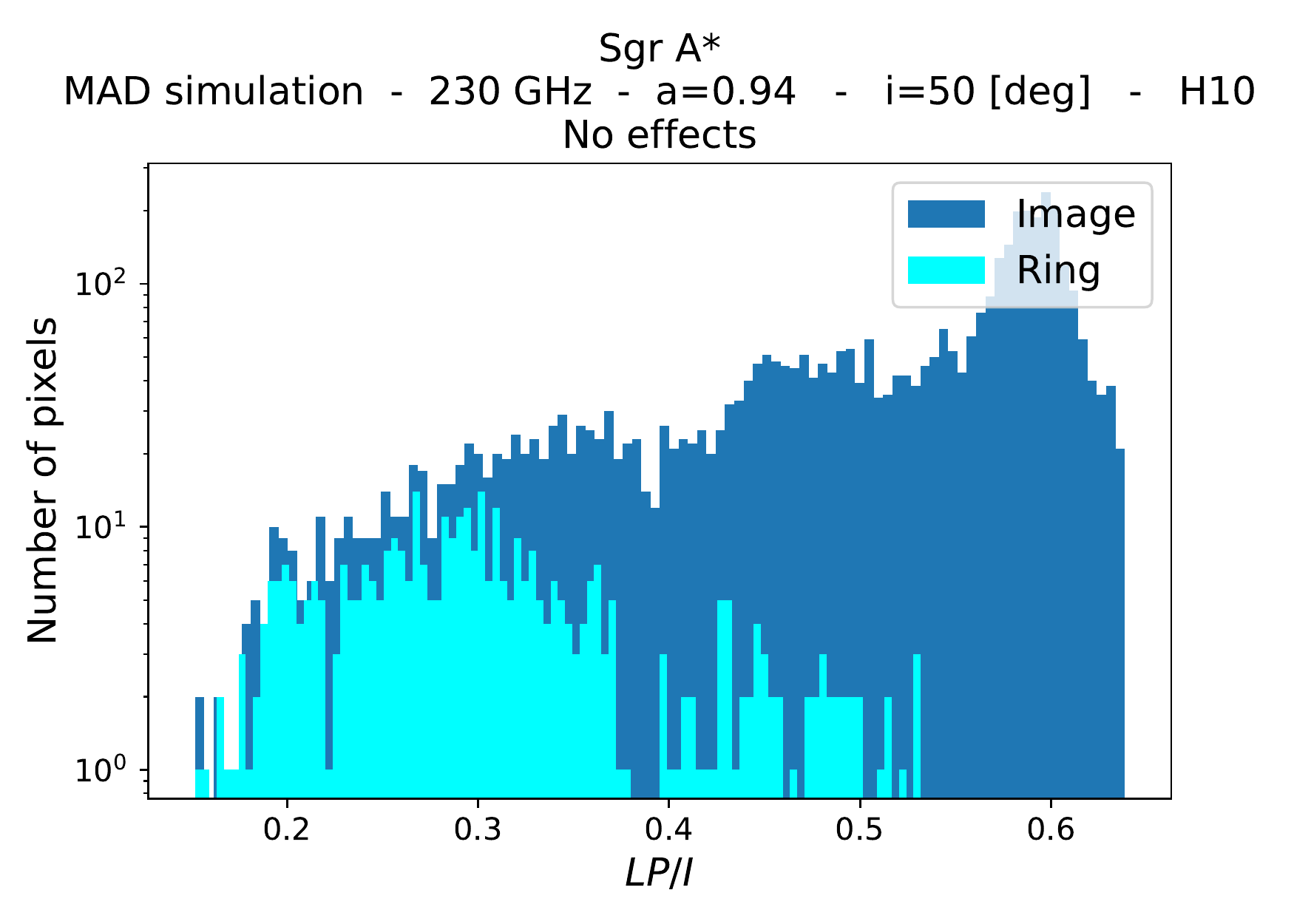}
\includegraphics[trim = 0cm 0cm 0cm 1.55cm, clip=true, width=0.45\textwidth]{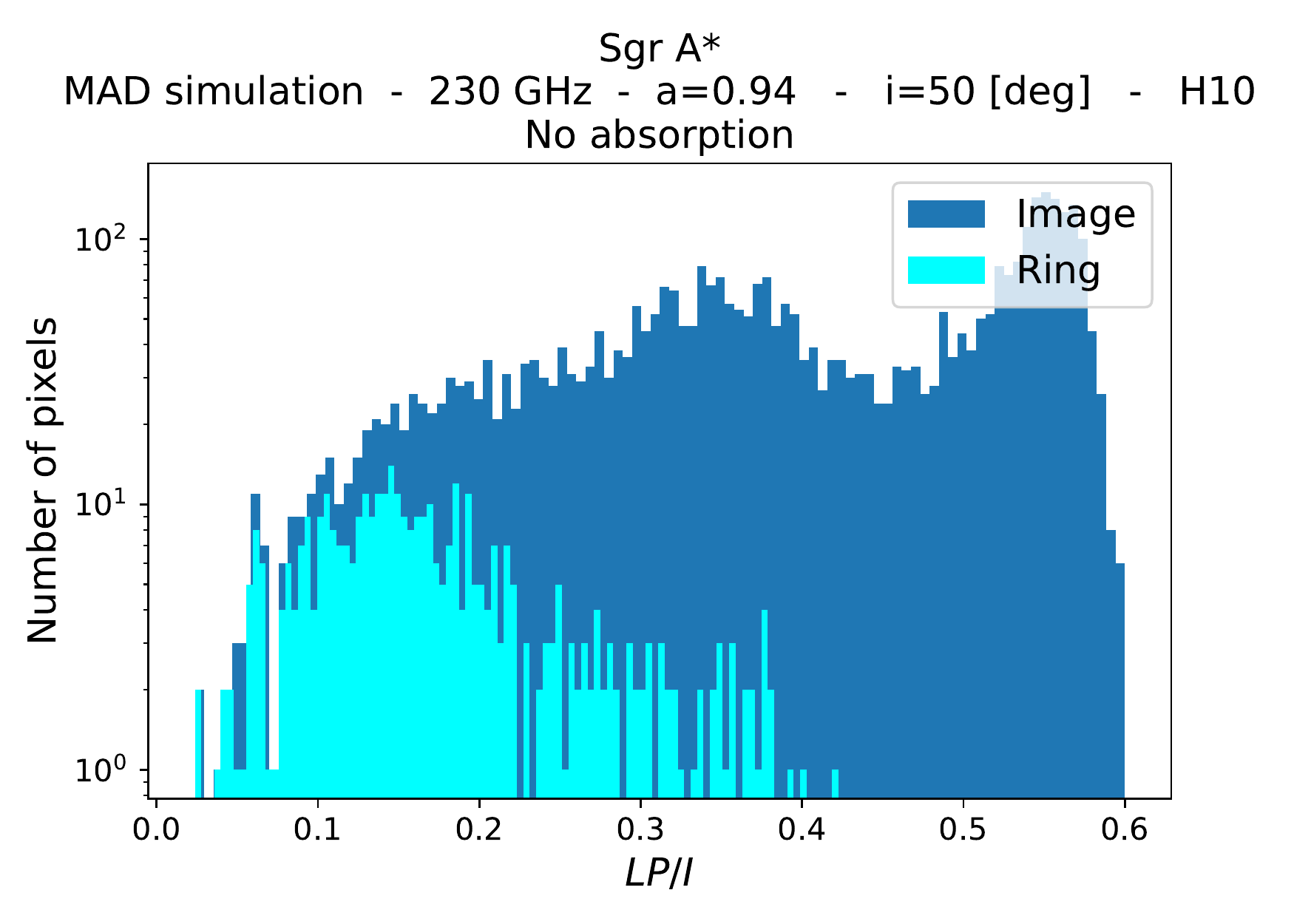}
\includegraphics[trim = 0cm 0cm 0cm 1.55cm, clip=true, width=0.45\textwidth]{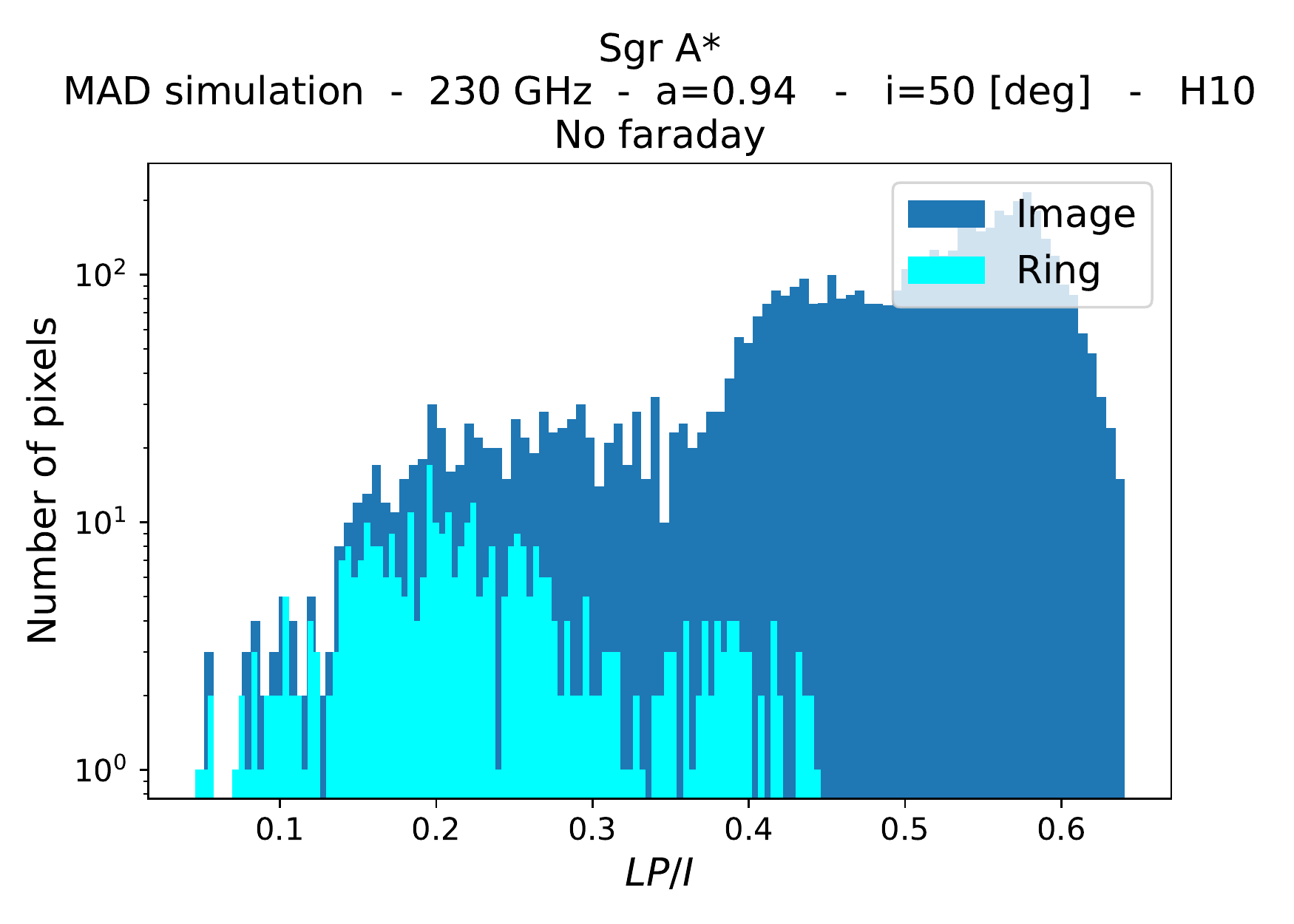}
 \caption{Distribution of image and ring fractional polarized flux ($LP/I$) per pixel for the MAD simulation. Only pixels with a total intensity of at least $10\%\ I_{\rm{med}}$ are considered, with $I_{\rm{med}}$ the median of the one hundred brightest pixels. Image pixels are in dark blue, ring pixels are in cyan. $LP/I$ is calculated from thirty-frame averaged images where different effects are considered. Top left: all effects. Top right: no effects. Bottom left: no absorption. Bottom right: no Faraday effects. In all cases the photon ring pixels are preferentially less polarized relative to the rest of the image.}
     \label{fig:histograms_MAD}
\end{figure*}

\begin{figure*}
\centering
\includegraphics[trim = 3cm 8.8cm 3cm 0cm, clip=true,width=1\textwidth]{./figures_final/sgra_sane_gmin3_all_effects_230i50a0_panels.pdf}
\includegraphics[trim = 0cm 0cm 0cm 1.55cm, clip=true, width=0.45\textwidth]{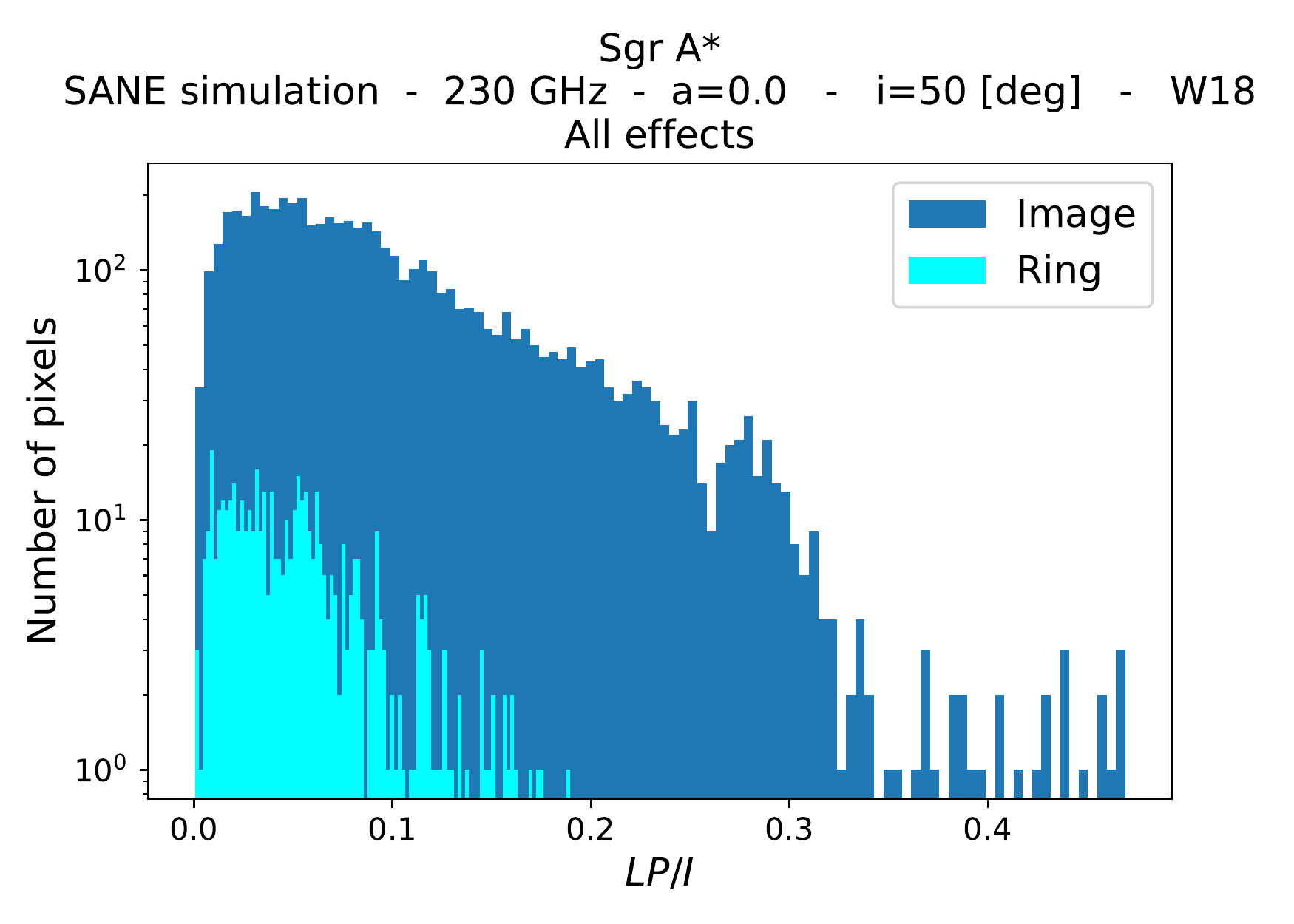}
\includegraphics[trim = 0cm 0cm 0cm 1.55cm, clip=true, width=0.45\textwidth]{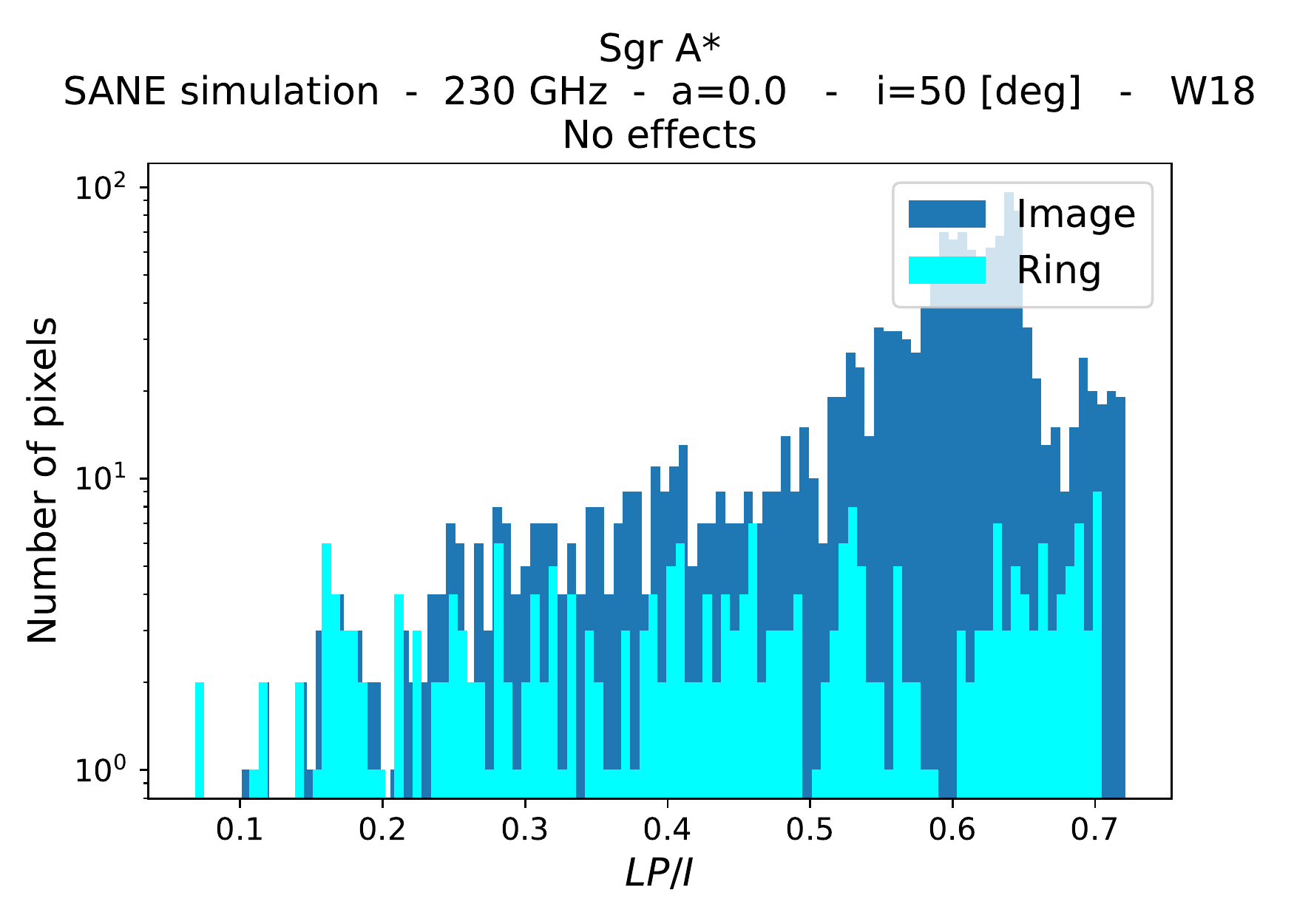}
\includegraphics[trim = 0cm 0cm 0cm 1.55cm, clip=true, width=0.45\textwidth]{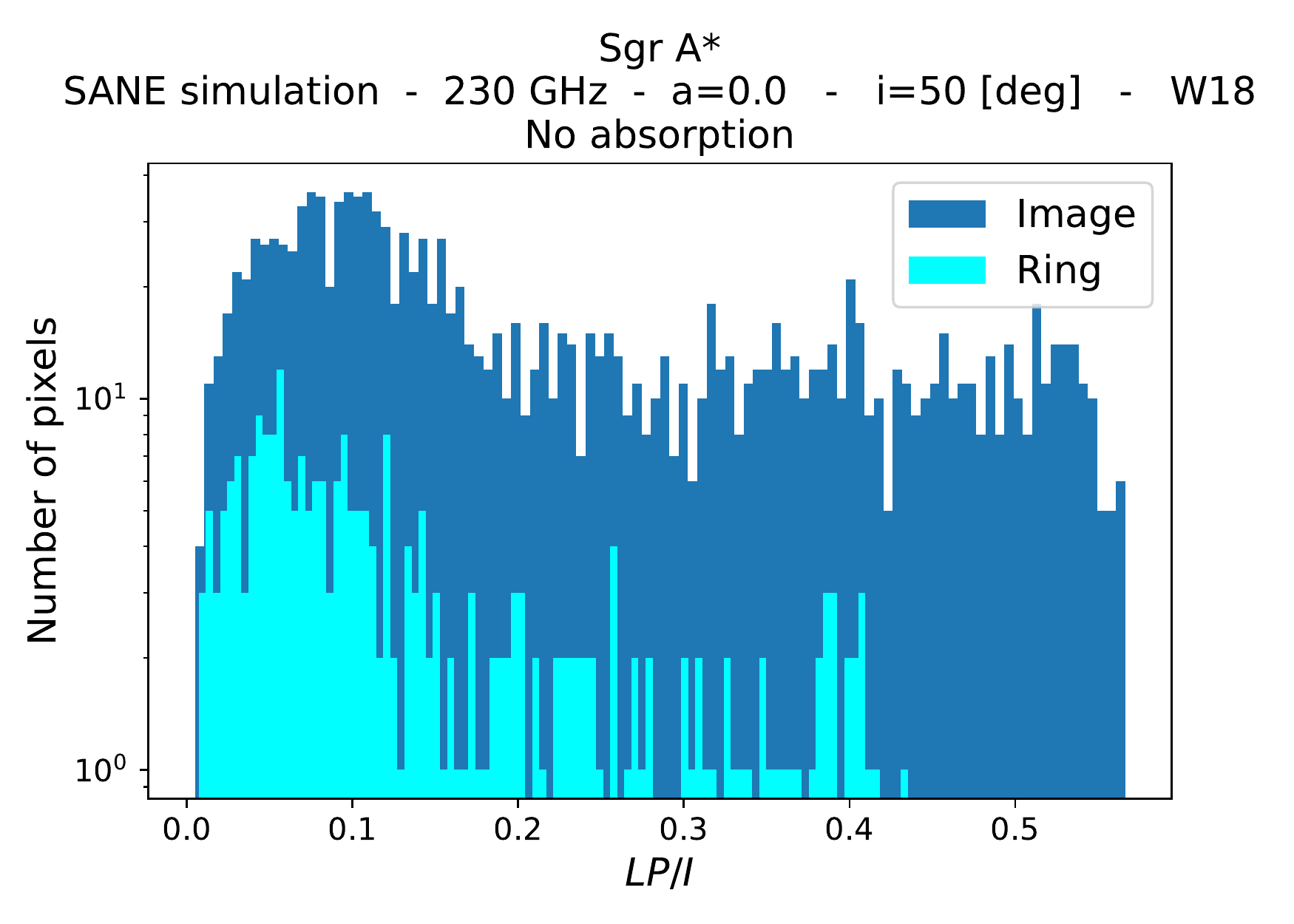}
\includegraphics[trim = 0cm 0cm 0cm 1.55cm, clip=true, width=0.45\textwidth]{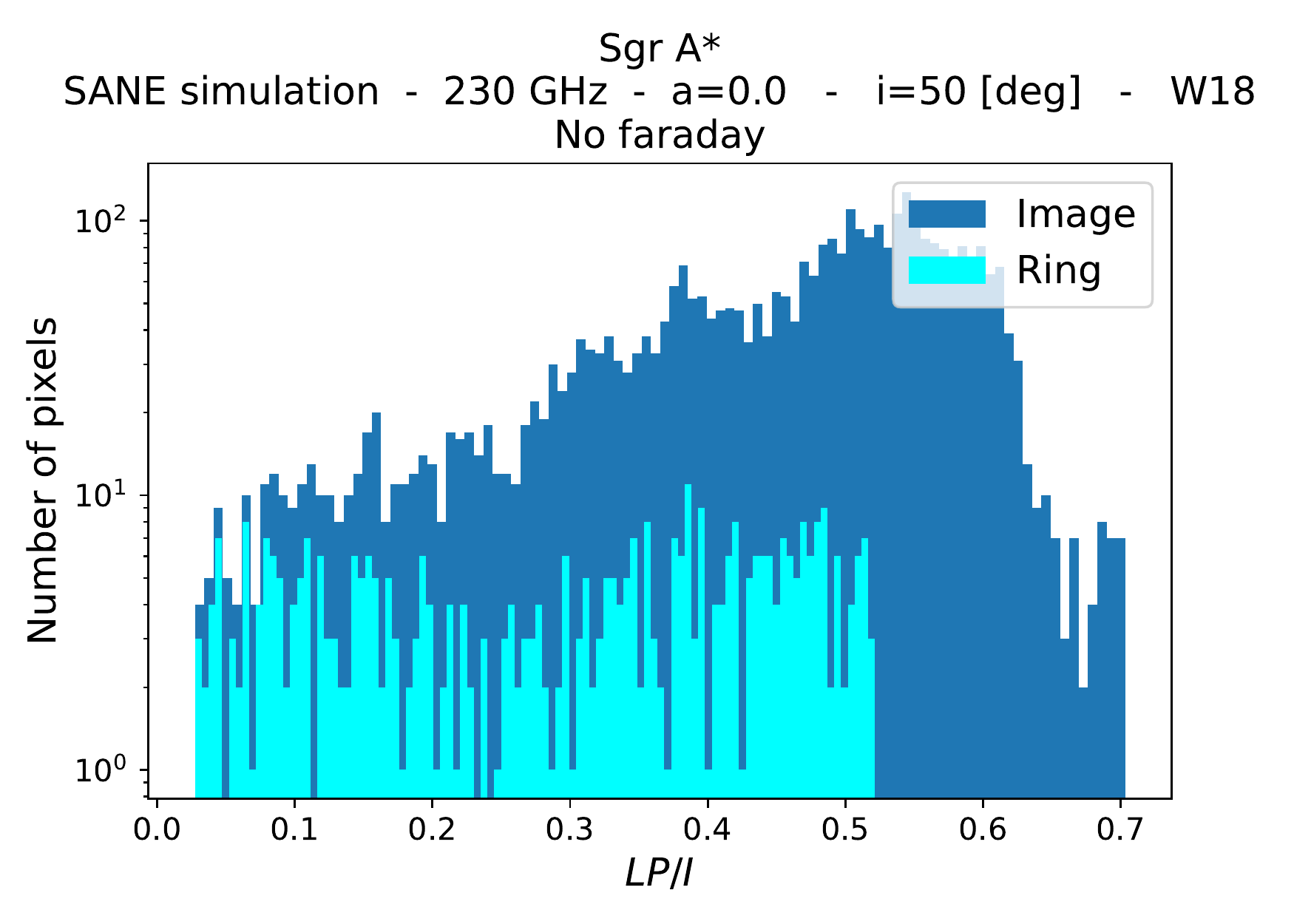}
 \caption{Distribution of $LP/I$ per pixel for the SANE simulation. Only pixels with a total intensity of at least $10\%\ I_{\rm{med}}$ are considered, with $I_{\rm{med}}$ the median of the one hundred brightest pixels. The values are calculated from thirty-frame averaged images where different effects are considered. Image pixels are in dark blue, ring pixels are in cyan. Top left: all effects. Top right: no effects. Bottom left: no absorption. Bottom right: no Faraday effects. Just like in the MAD case, the photon ring pixels are clear outliers in the distribution.}
     \label{fig:histograms_SANE}
\end{figure*}

We investigate the reasons for which the photon ring has different properties from the rest of the image. To do this, we study how the polarization properties of the photon ring are affected in the presence/absence of effects that can depolarize an image.
We consider four effects: absorption, Faraday rotation and conversion, magnetic turbulence and parallel transport.


\subsection{Effects of absorption, Faraday rotation and Faraday conversion}

\textsc{grtrans} calculates images by solving the polarized radiative transfer equation along geodesics traced between where a photon was emitted and the observer's camera. By setting the absorption, Faraday rotation or conversion coefficients in the radiative transfer equation to zero, we can study the influence of these effects straightforwardly. 

We first look at the distribution of $LP/I$ in the image and photon ring pixels from a set of images where either all (or none) of the effects have been taken into account. In order to avoid biases from dim regions in the image, only pixels with a total intensity of at least $10\%\ I_{\rm{med}}$ have been considered, with $I_{\rm{med}}$ the median of the one hundred brightest pixels. 

Figures \ref{fig:histograms_MAD} and \ref{fig:histograms_SANE} show that the photon ring is a clear outlier in the distribution of polarized flux in all cases. When absorption and/or Faraday rotation are neglected, the polarization fraction increases everywhere. Still the photon ring remains fairly less polarized compared to the rest of the image.

The net fractional LP in the ring and image is given by $m_x = \langle \rm{LP/\ I} \rangle$ where $x$ denotes whether the calculation uses image or photon ring pixels. The values of $m$ for the MAD and SANE cases are shown in Tables \ref{tab:fluxes_MAD_med} and \ref{tab:fluxes_SANE_med}. In all cases the photon ring is about a factor of $\simeq 2-2.5$ less polarized than the rest of the image, with maximum depolarization obtained when all effects are included in the calculations.
As expected, the highest degree of polarization in the ring ($\sim60-75\%$ of that of the image) is achieved when neither effect is included in the calculation. 
Absorption has a larger impact on the depolarization degree of the ring than Faraday effects (about $\sim10\%$ more). However, neither effect alone (or combined) amounts to the total observed level of depolarization in the photon ring. A comparison between the ``all effects'' and ``no effects'' cases shows that for these simulations the plasma effects decrease the relative fractional polarization in the ring, $m_{\rm{Ring}}/m_{\rm{Image}}$, by $\sim20\%$, while other effects do so by $\sim30\%$.

\begin{table}
  \centering
	\caption{Image and ring net fractional LP ($m$) of the MAD simulation. The values are calculated from a thirty-frame averaged image including effects of absorption and/or Faraday rotation and conversion at a time. 
	}
  \label{tab:fluxes_MAD_med}
  \resizebox{\columnwidth}{!}{%
  \begin{tabular}{*{6}{|c}|}
  \hline
  & & All effects & No effects & No absorption & No Faraday \\
  \hline\hline
  MAD & $m_{\rm{Image}}\ [\%]$ & 44.66 & 51.89 & 47.88 & 48.23 	\\
	  & $m_{\rm{Ring}}\ [\%]$  & 20.73 & 32.45 & 27.46 & 23.94  \\
      & $m_{\rm{Ring}}\ /\ m\ _{\rm{Image}} $ & 0.46  & 0.63  & 0.57  & 0.50 \\
  \hline\hline
\end{tabular}%
}
\end{table}

\begin{table}
	\centering
	\caption{Image and ring net fractional LP ($m$) of a thirty-frame averaged SANE image with different effects included in the calculations. 
	}
	\label{tab:fluxes_SANE_med}
	\resizebox{\columnwidth}{!}{%
  \begin{tabular}{*{6}{|c}|}
		\hline
		\hline
		 & & All effects & No effects & No absorption & No Faraday \\
    	\hline
		SANE & $m_{\rm{Image}}\ [\%]$ & 30.87 & 53.69 & 48.56 & 43.79 \\
		     & $m_{\rm{Ring}} \ [\%]$ & 18.00 & 39.76 & 35.77 & 25.94 \\
     & $m_{\rm{Ring}}\ /\ m_{\rm{Image}}$ & 0.58  & 0.74  & 0.74  & 0.59 \\
		\hline
		\hline
	\end{tabular}
	}
\end{table}


\subsection{Effects of magnetic turbulence and parallel transport}

\begin{figure*}
\centering
\includegraphics[trim = 0cm 8.8cm 0cm 0cm, clip=true,width=1\textwidth]{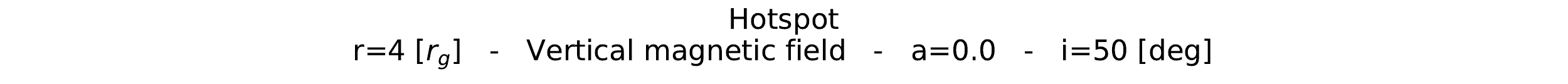}
\includegraphics[trim = 0.5cm 0cm 0cm 1.55cm, clip=true,width=0.65\textwidth]{./figures_final/hts_r4_vertical_a0_panels.pdf}
\includegraphics[trim = 0cm 0cm 1cm 1.6cm, clip=true,width=0.34\textwidth]{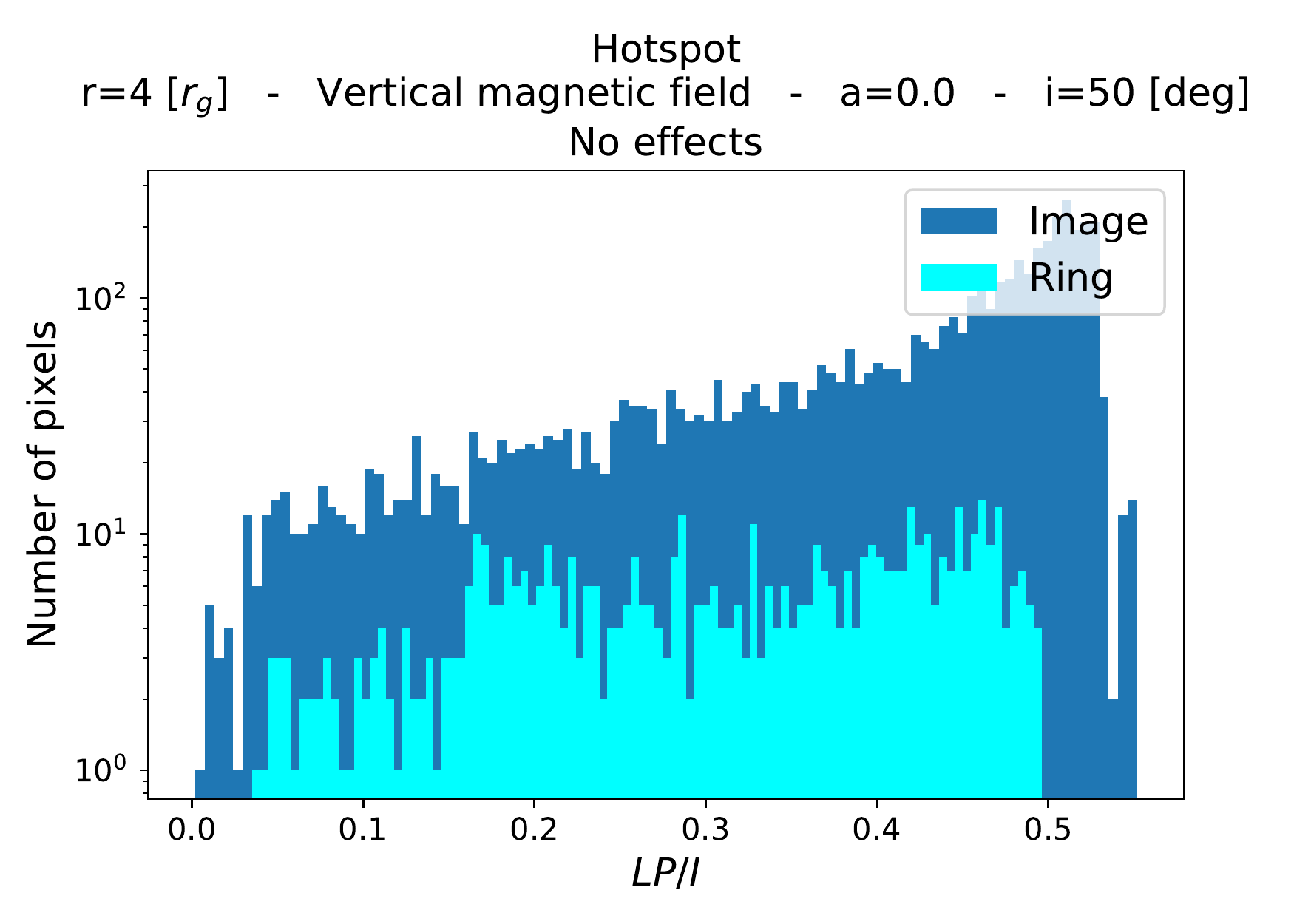}
 \caption{Stacked frames of a hotspot orbiting the black hole with spin $a=0.0$. The hotspot is at a distance $4\ r_g$, traces clockwise motion and moves in a completely vertical magnetic field. The observer's inclination angle is $i=50^\circ$. In this simplified model the magnetic field is fully ordered and both self-absorption and Faraday rotation are negligible. 1) Total intensity. 2) Polarized flux. 3) Residual image. 4) Distribution of $LP/I$ of the image and photon ring pixels. Only pixels with a total intensity of at least $10\%\ I_{\rm{med}}$ are considered, with $I_{\rm{med}}$ the median of the one hundred brightest pixels. Image pixels are in dark blue, ring pixels are in cyan. Unlike the MAD and SANE cases, the photon ring pixels follow a similar distribution compared to the rest of the image. }
\label{fig:snapshots_hotspot}
\end{figure*}

We next look into the effects of magnetic field turbulence and gravitational effects.
We expect magnetic field turbulence and parallel transport to depolarize emission in the ring due to multiple contributions to the polarized intensities $j_{Q,U}$ along the light rays (which wrap around the black hole), which will tend to average down.
However, while parallel transport is constant in time and model independent, magnetic field structure can be vastly different in models and exhibit time dependent behaviour. 


In order to study turbulent magnetic field structure effects more, we compare our MAD and SANE GRMHD results to idealized semi-analytical models with an ordered magnetic configuration and no radiative transfer effects. We consider an optically-thin compact emission region \citep[``hotspot'', ][]{broderick2006} orbiting a black hole. 
The hotspot is moving with constant speed (matching that of a test particle at its center) in the equatorial plane at orbital radius $R_0$.
We consider two magnetic field geometries, vertical and toroidal; two different orbital radii, $4\ r_g$ and $8\ r_g$; and three black hole spins, $a=-0.9,0.0,0.9$. 
The maximum particle density $n_{\rm spot}\sim2\times10^7 \, \rm cm^{-3}$ falls off as a 3D Gaussian with characteristic size $R_{\rm spot}$. We fix the viewer's inclination at $i=50$ degrees, consistent with GRAVITY flare observations \citep{gravity2018flare,gravity2020_michi,gravity2020_pol}.
We calculate synchrotron radiation from a power law distribution of electrons with a minimum Lorentz factor of $1.5\times10^3$. 

Example images of total intensity and polarized flux of one of the models are presented in panels 1 and 2 of Fig.~\ref{fig:snapshots_hotspot}. The panels show stacked images over one revolution of the hotspot. The orbital radius is $R_0=4 \ r_g$, the black hole spin $a=0.0$ and the magnetic field geometry is completely vertical. Since the emission is subject to no effects other than magnetic field structure and gravitational effects, there is a lack of dark depolarized zones in the $LP$ image. We can see gravitational lensing and beaming in the form of secondary images, asymmetry in the emission (top brighter than bottom due to Doppler beaming from the hotspot approaching the observer) and the photon ring.

\begin{table}
  \centering
	\caption{Net fractional LP ($m$) of stacked images over one revolution of a hotspot model in different magnetic field (B) geometries. 
	}
	\label{tab:fluxes_hotspot_all_med}
	\begin{tabular}{*{6}{|c}|}
   \hline
     Hotspot & B field & Spin & $m_{\rm Image}\ [\%]$ & $m_{\rm Ring}\ [\%]$& $m_{\rm{Ring}}\ /\ m_{\rm Image}$ \\
    \hline\hline
		 $4 \ r_g$ & Vertical & -0.9  &  41.10  &  29.72  & 0.72  \\
	               &          & 0.0   &  40.74  &  33.17  & 0.81  \\
	               &          & 0.9   &  43.53  &  36.56  & 0.84  \\
		 $8 \ r_g$ & Vertical & -0.9  &  51.48  &  48.13  & 0.93  \\
                   &          & 0.0   &  51.83  &  47.94  & 0.92  \\
                   &          & 0.9   &  52.02  &  47.02  & 0.90  \\
        \hline
		 $4 \ r_g$ & Toroidal & -0.9  &  43.13  &  41.97  & 0.97  \\
	               &          & 0.0   &  41.83  &  43.02  & 1.03  \\
	               &          & 0.9   &  39.28  &  40.56  & 1.03  \\
		 $8 \ r_g$ & Toroidal & -0.9  &  48.33  &  48.47  & 1.00  \\
                   &          & 0.0   &  47.70  &  48.58  & 1.02  \\
                   &          & 0.9   &  47.07  &  47.80  & 1.02  \\
    \hline\hline
    \end{tabular}%
\end{table}

The distribution of fractional $LP$ per pixel in the idealized model (right panel of Fig.~\ref{fig:snapshots_hotspot}) shows that the photon ring pixels have a similar behaviour to that of the rest of the image. This is supported by the $m$ values (Table \ref{tab:fluxes_hotspot_all_med}). Similar results are found for the rest of the models.  

In contrast to the MAD and SANE cases, $m_{\rm Ring}\simeq m_{\rm Image}$ in the hotspot models, which suggests that complex magnetic field structure is an important factor for depolarizing the photon ring.

Tables \ref{tab:fluxes_MAD_med} and \ref{tab:fluxes_SANE_med} show that generally speaking, the SANE simulation is more depolarized than the MAD (up to $\sim60\%$ depolarization with respect to the rest of the image when all effects are considered).
Comparing both ``no effects'' cases for the MAD and SANE simulation, where no radiative transfer effects are included in the calculations, we see that their net polarized fluxes have very similar values. In the SANE case the weak fields are sheared into a predominantly toroidal configuration in the accretion flow, while in the MAD case significant poloidal components remain. 
While complex or turbulent magnetic field structure appears to be an important contributor to depolarization of the photon ring, the effect is not strongly model-dependent, e.g., the details appear to be less important.

In the case of the hotspot, Table~\ref{tab:fluxes_hotspot_all_med} shows that for the spin values and orbital radii considered, we do not find strong depolarization in the photon ring compared to the rest of the image $m_{\rm{Ring}}/m_{\rm{Image}} \simeq 0.85$.


\section{Summary and discussion}
\label{sec:summary}

\begin{figure*}
\centering
\includegraphics[trim = 0cm 0cm 0cm 0cm, clip=true, width=1\textwidth]{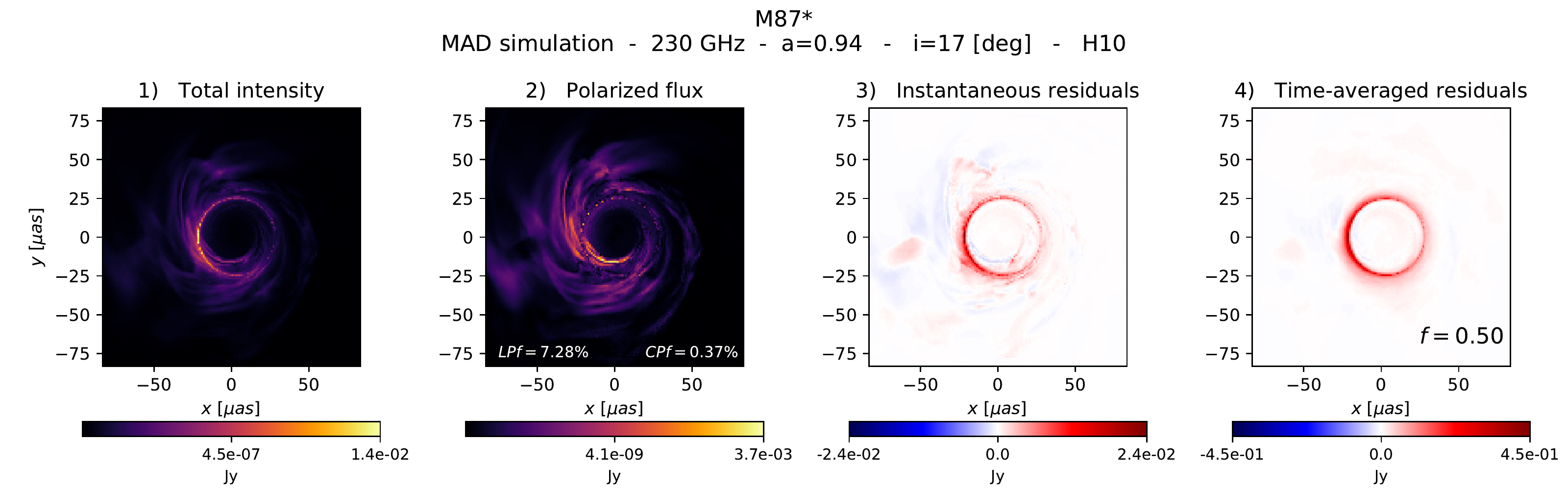}
\includegraphics[trim = 0cm 0cm 0cm 0cm, clip=true, width=1\textwidth]{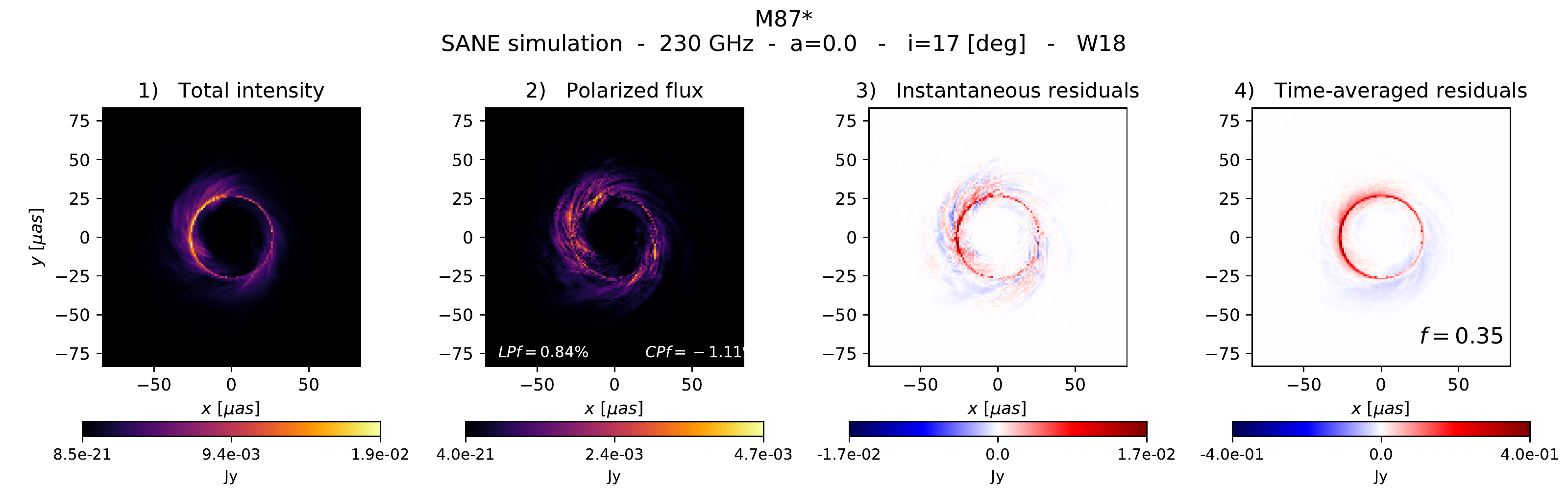}
 \caption{Photon extraction technique applied to 230 GHz images of M87*. The images are calculated from the same GRMHD simulations used for Sgr A*. The black hole mass is scaled to $6\times10^9\ \rm{M_\odot}$ and the total flux to $0.5-1$ Jy. The viewer's inclination is $i=17^\circ$. Total intensity (1), polarized flux (2), instantaneous residuals (3) and time-averaged residuals (4). Top: MAD simulation. Bottom: SANE simulation. In both cases the technique works well for extracting the photon ring. 
 }
     \label{fig:m87_panels}
\end{figure*}
\begin{figure*}
\centering
\includegraphics[trim = 0cm 0cm 0cm 0cm, clip=true, width=1\textwidth]{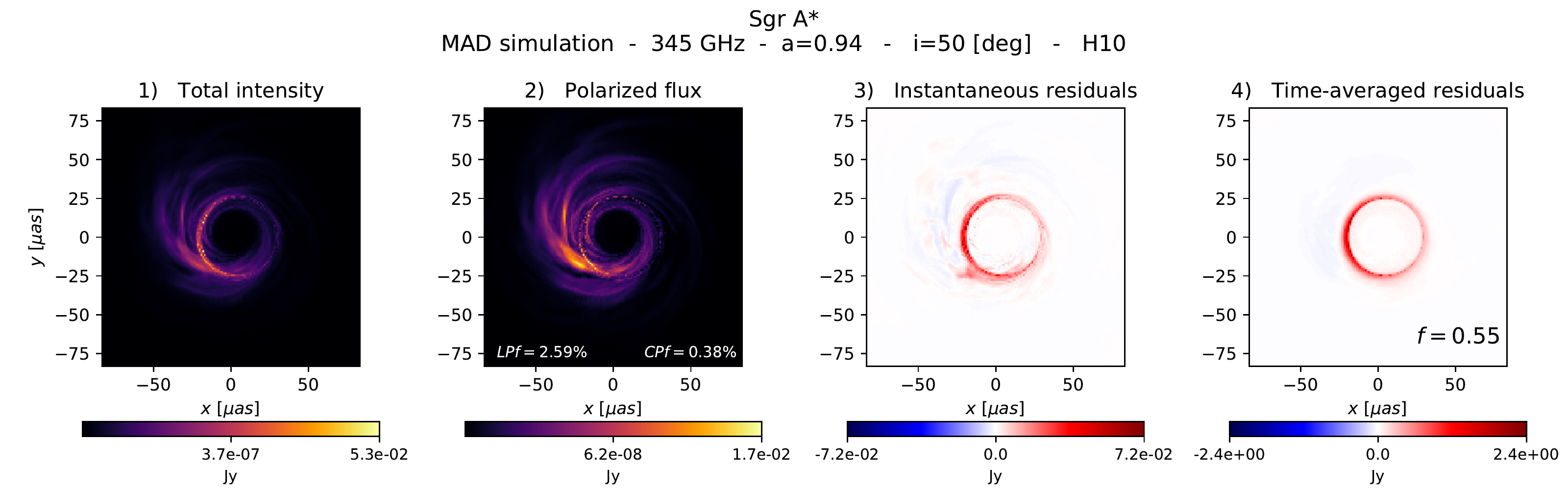}
 \caption{Total intensity (1), polarized flux (2), instantaneous residuals (3) and time-averaged residuals (4) with $f=0.55$ of images at $345$ GHz of the long duration, MAD GRMHD simulation. The net LPf and CPf in the image are indicated in Panel 2. The extraction of the photon ring works better at higher frequency, where other depolarizing effects of self-absorption and Faraday rotation are weaker.
 }
     \label{fig:snapshots_MAD_345GHz}
\end{figure*}

\begin{figure*}
\centering
\includegraphics[trim = 0cm 0cm 0cm 0cm, clip=true,width=1\textwidth]{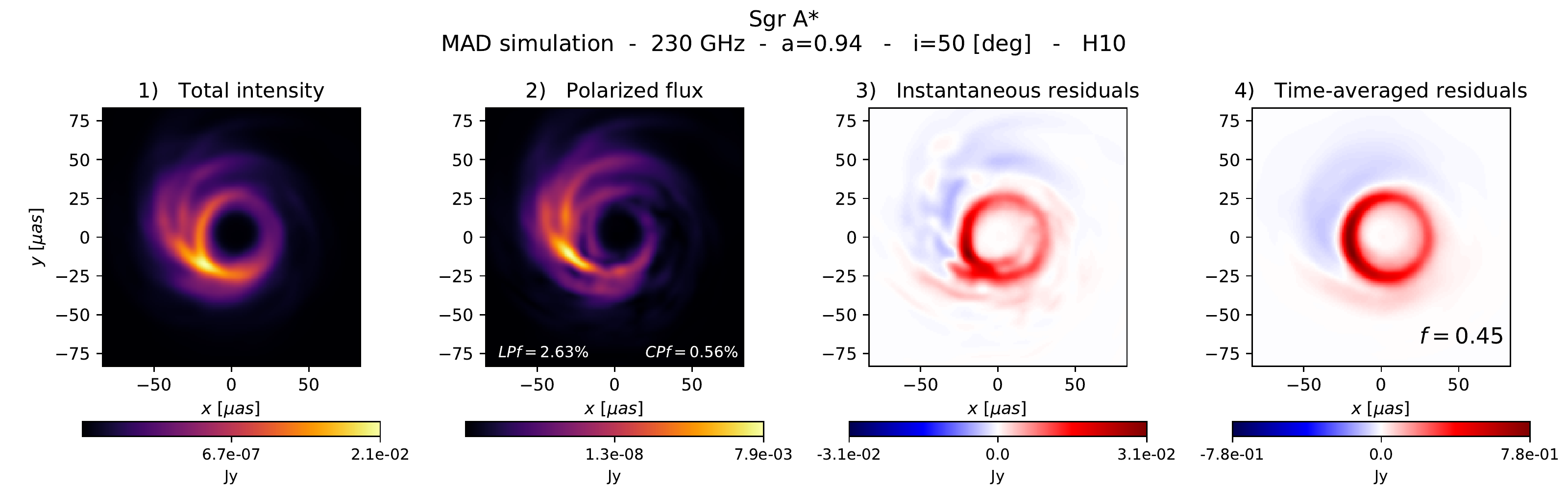}
\includegraphics[trim = 0cm 0cm 0cm 0.65cm, clip=true, width=1\textwidth]{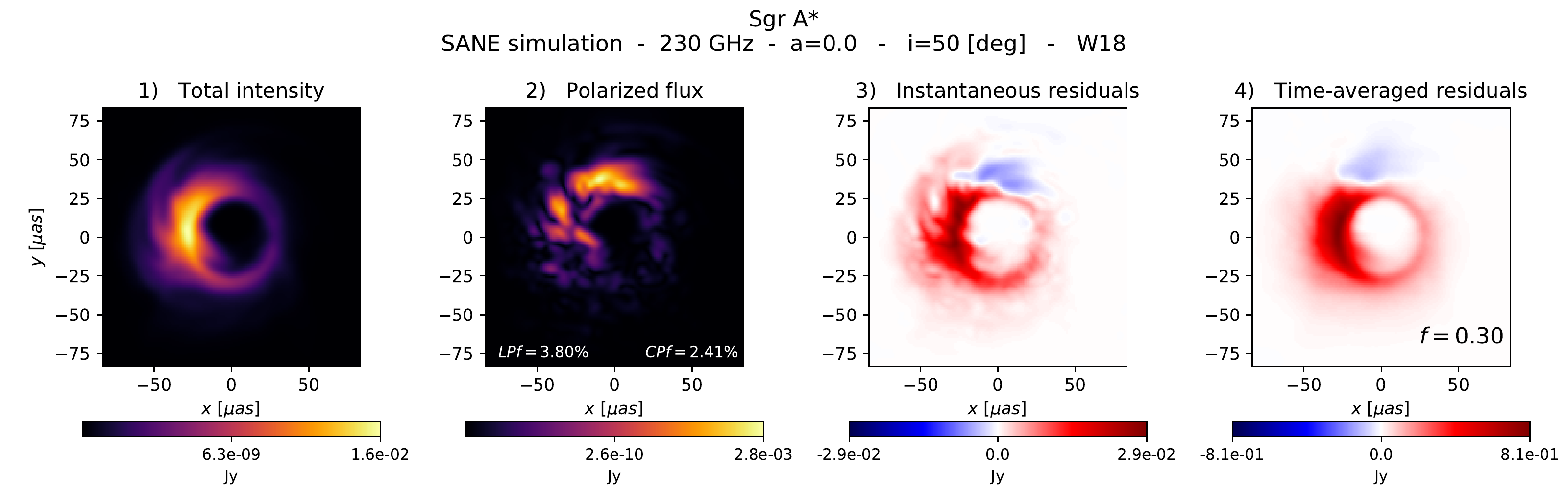}
 \caption{ Photon ring extraction technique applied to GRMHD images blurred with a $20\mu$as Gaussian. Top row: MAD simulation. Bottom row: SANE simulation. (1) Total intensity. (2) Polarized flux. (3) Instantaneous residuals. (4) Time-averaged residuals over thirty frames. The technique is still effective for the MAD case. In the SANE case however, the polarized flux image does not show a prominent ring and it is unclear if the ring features in time-averaged residuals are better than the total intensity image.  }
     \label{fig:blurred_GRMHD}
\end{figure*}

We have shown that an optimal subtraction of linear polarized flux images from total intensity images calculated from GRMHD simulations can enhance the photon ring feature when the conditions in the plasma are suitable. 
The method relies on the fact that both kinds of images are highly correlated, except at the photon ring. The photon ring is preferentially less polarized with respect to the rest of the image.

The latter implies that time variability in the source favours a better extraction of the ring. If the photon ring is the only persistently less polarized image feature, residuals from depolarized pixels in the rest of the image will be suppressed when time-averaging, sharpening the photon ring feature. 
Given that the black hole mass sets the length and timescales of the system, compared to Sgr A*, M87* is about a thousand times more massive and a thousand times slower. 
Time-averaging for Sgr A* could be done within a single day or epoch of observing, while for M87* it would require many observational campaigns spanning months to years.

Extracting the photon ring by subtracting polarized flux requires a prominent photon ring feature in the total intensity image, and fails when regions other than the photon ring are persistently depolarized, e.g., due to self-absorption or Faraday rotation. As a result, the method works best at low to moderate observer inclinations. M87* is known to be viewed at low inclination and viable theoretical models of the EHT image show prominent photon ring features \citep{gravity2018flare,gravity2020_michi,gravity2020_pol}. 
We present in Fig.~\ref{fig:m87_panels} the photon extraction technique applied to 230 GHz images of M87*.  These snapshots are calculated using the GRMHD simulations described in Section~\ref{sec:method}. The black hole mass has been scaled to $6\times10^9\ \rm{M_\odot}$ and the total flux at 230 GHz to $0.5-1$ Jy. The viewer’s inclination is $i=17^\circ$. It can be seen that the technique works well in these low inclination cases where the photon ring is visible.

Frequency is an important factor as well. At low frequencies where the emission is self-absorbed, and in local thermodynamic equilibrium, the intensity is that of a blackbody. The radiation will tend to be unpolarized not only at the photon ring, but anywhere in the accretion flow where the optical depth is large. In this case, the photon ring will no longer be an outlier and time variability will not improve the residuals image over time since the location of zones with high optical depth will not vary much. Faraday effects will affect the image in a similar manner, becoming stronger with decreasing frequency and introducing depolarization all over the image. Frequencies $\nu\gtrsim230$ GHz produce even more favourable results. In particular, Fig.~\ref{fig:snapshots_MAD_345GHz} shows a sharp ring residual image at 345 GHz for our MAD model.

We have looked into the numerical properties of the photon ring in our GRMHD images. The lower degree of linear polarization with respect to the rest of the image ($\sim 2-2.5$ less polarized) is  caused by a combination of plasma effects (absorption and Faraday effects), gravitational effects and magnetic field turbulence in the plasma. 
The method appears to work as well for sample snapshot polarized images (C. J. White, private communication) from \citet[][]{ressler2020mad}, where the simulations are initialized from larger scale MHD simulations of accretion onto Sgr A* via magnetized stellar winds. Those simulations show a complex magnetic field structure, where the magneto rotational instabilities are not important at any time \citep{ressler2020}. 
We explore other models in Appendix~\ref{appendix:other_models} and find that the photon ring continues to be less polarized than the rest of the image. An interesting case is shown in Appendix~\ref{appendix:MADa0}, where $m_{\rm{Ring}}\lesssim m_{\rm{Image}}$, even when all plasma effects are included in the calculations. Understanding when depolarization of the photon ring is seen in GRMHD is an important goal for future work.

We have compared our GRMHD results to those of a semi-analytical hotspot model in idealized magnetic field configurations, different orbital radii and black hole spin. In this case, the results seem consistent with analytic predictions for optically thin emission around black holes \citep{himwich2020}, where parallel transport is not expected to depolarize emission in the photon ring. 
Evidently, in addition to plasma effects, both parallel transport and some degree of disorder in the magnetic field configuration are required to cause the depolarization of the photon ring that we find. The results appear insensitive to the model chosen, among the few we have explored.

So far the technique has been presented for the image domain only, using an angular resolution of $\simeq 1 \mu$as, far superior to that available with the EHT ($\simeq 10-20 \mu$as). Given the non-linearity of the polarized flux in terms of the observable Stokes parameters $Q$ and $U$, an extension to the Fourier domain, where the observables are measured, is non-trivial and is left for future work.

We tried blurring images by convolving the map of each Stokes parameter with a $20\mu$as FWHM Gaussian kernel and then applying the technique. The results are shown in Figure \ref{fig:blurred_GRMHD}. Whereas for the MAD scenario the method still appears to be promising, resulting in a prominent blurred ring, in the SANE case the ring features in the time-averaged residual map do not appear much clearer than in the original total intensity image.

For conditions similar to those found in many viable GRMHD models of Sgr A* and M87*, the relative depolarization of emission from the black hole photon ring may provide a powerful method for extracting it from horizon scale images and movies. The method naturally benefits from the turbulent variability expected from accreting black holes.

\section*{Acknowledgements}
We thank Chris J. White for providing sample images. J.D. and A.J.-R. were supported in part by a Sofja Kovalevskaja award from the Alexander von Humboldt foundation, by a CONACyT/DAAD grant (57265507), and by NASA Astrophysics Theory Program Grant 80NSSC20K0527. S.M.R. is supported by the Gordon and Betty Moore Foundation through Grant GBMF7392. AT was supported by the National Science Foundation AAG grants 1815304 and 1911080. The calculations presented here were carried out on the MPG supercomputers Hydra and Cobra hosted at MPCDF.

\section*{Data Availability}
The simulated images used here will be shared on reasonable request to the corresponding author.




\bibliographystyle{mnras}
\bibliography{references} 



\appendix
\section{Dependence of results on ray tracing image resolution}
\label{appendix:resolution_effects}

\begin{figure*}
\centering
\includegraphics[trim = 3cm 8.8cm 3cm 0cm, clip=true,width=1\textwidth]{./figures_final/sgra_mad_gmin1_all_effects_230i50a94_panels.pdf}
\includegraphics[trim = 0cm 0cm 0cm 0cm, clip=true,width=0.45\textwidth]{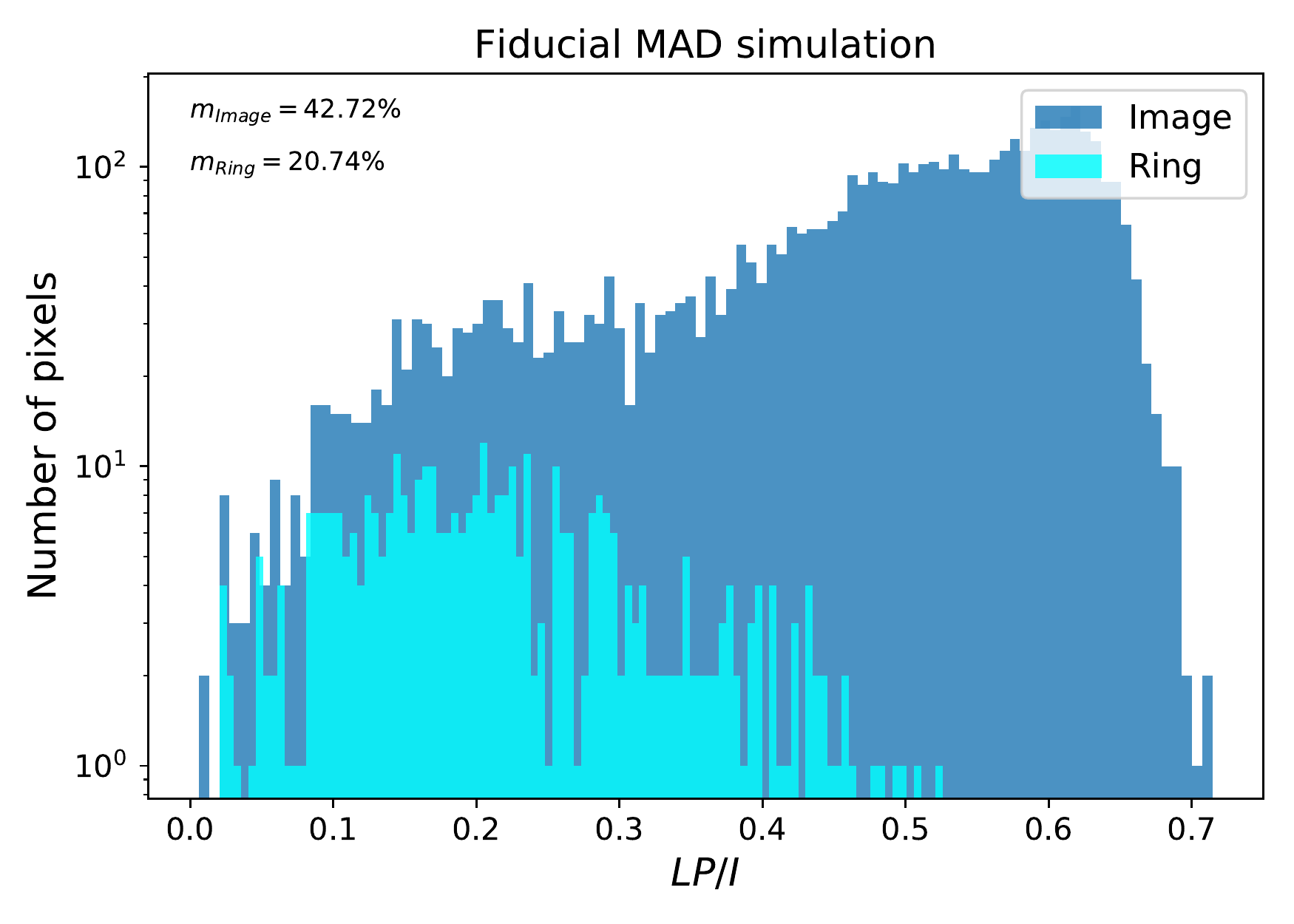}
\includegraphics[trim = 0cm 0cm 0cm 0cm, clip=true,width=0.45\textwidth]{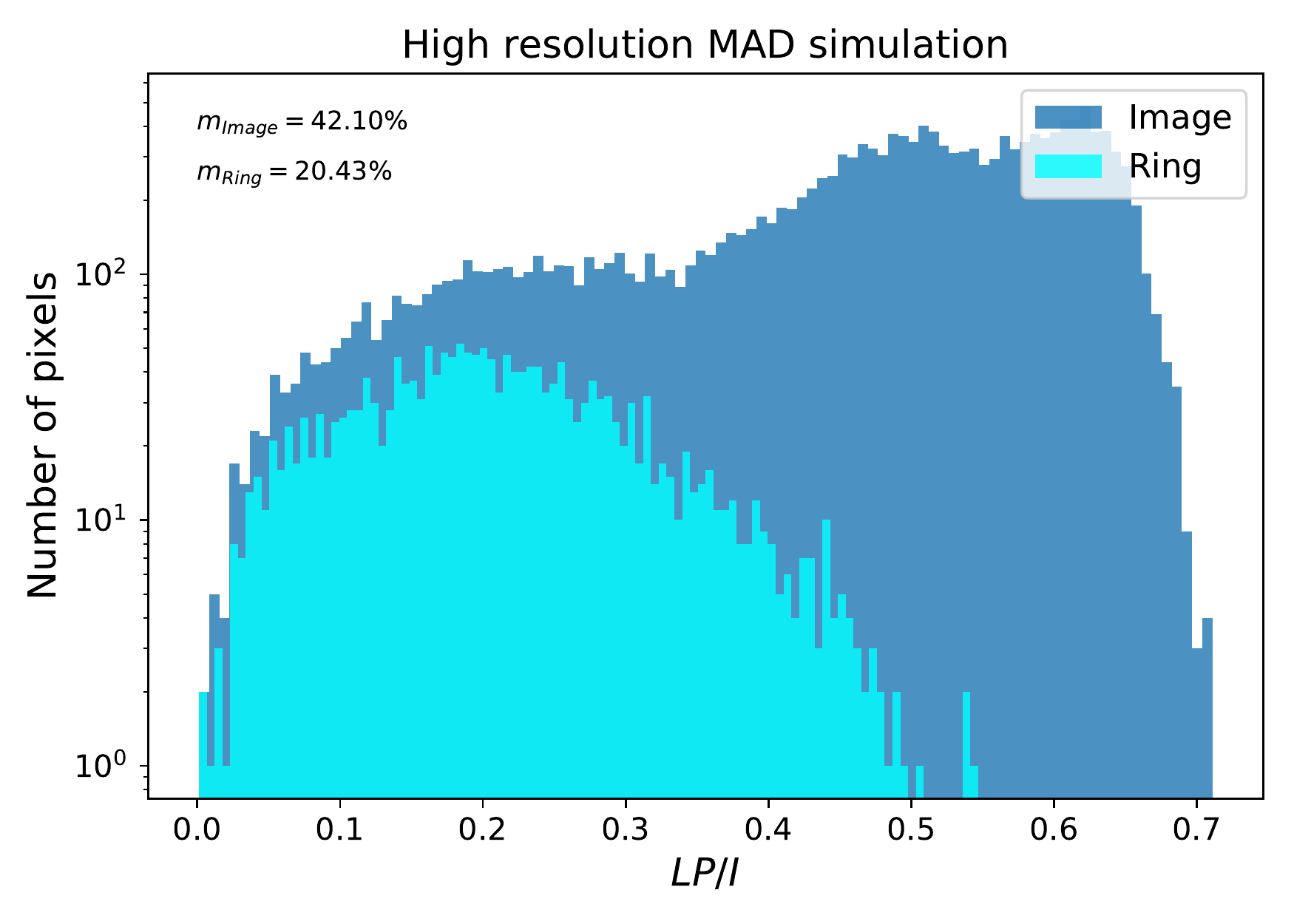}
 \caption{Distribution of $LP/I$ per pixel from one snapshot of the MAD simulation at different resolutions. All plasma effects have been considered in the calculation. Only pixels with a total intensity of at least $10\%\ I_{\rm{med}}$ are considered, with $I_{\rm{med}}$ the median of the one hundred brightest pixels. 
 Left: fiducial MAD simulation used in the calculations ($192\times192\times1600$ pixels, where the latter represents the step size of the geodesic integrator). 
 Right: higher  resolution ($384\times384\times1600$). 
 The net LP of the image and the ring are indicated on each panel. 
 Our results remain qualitatively the same with varying resolution: the photon ring pixels show different properties to the image pixels and are a factor $\sim 2-2.5$ less polarized.  }
\label{fig:histograms_resolution}
\end{figure*}

We have studied how image resolution affects the results presented in this work. Figure \ref{fig:histograms_resolution} shows distributions of the pixel fractional polarization from one snapshot of the MAD simulation at 230 GHz, at different resolutions. 
Though there are small differences among the exact values of the net LP, the photon ring pixels show different properties to those of the image pixels, with the photon ring being a factor of $\sim2-2.5$ less polarized. Our results remain qualitatively the same with resolution.


\section{Photon ring extraction method in other models}
\label{appendix:other_models}

We apply our extraction method to different models. 

\subsection{Different electron heating}

We calculate Sgr A* images at 230 GHz using the same MAD and SANE GRMHD simulations described in Section~\ref{sec:method}, but consider different electron heating mechanisms: MAD with magnetic reconnection (W18) and SANE with turbulent heating (H10). 

\begin{figure*}
\centering
\includegraphics[trim = 4cm 9.52cm 4cm 0cm, clip=true,width=1\textwidth]{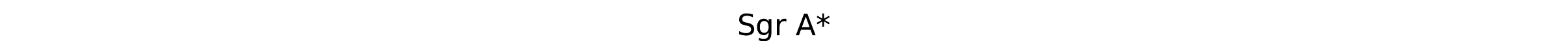}
\includegraphics[trim = 0cm 0cm 0cm 0.65cm, clip=true,width=1\textwidth]{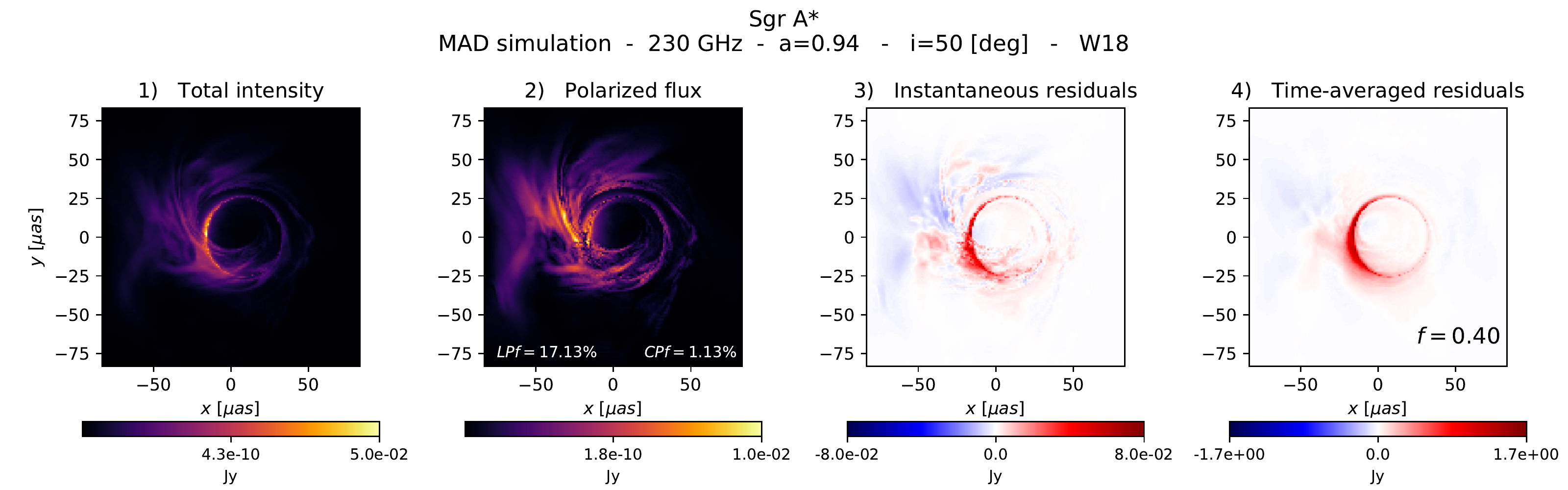}
\includegraphics[trim = 0cm 0cm 0cm 0.65cm, clip=true,width=1\textwidth]{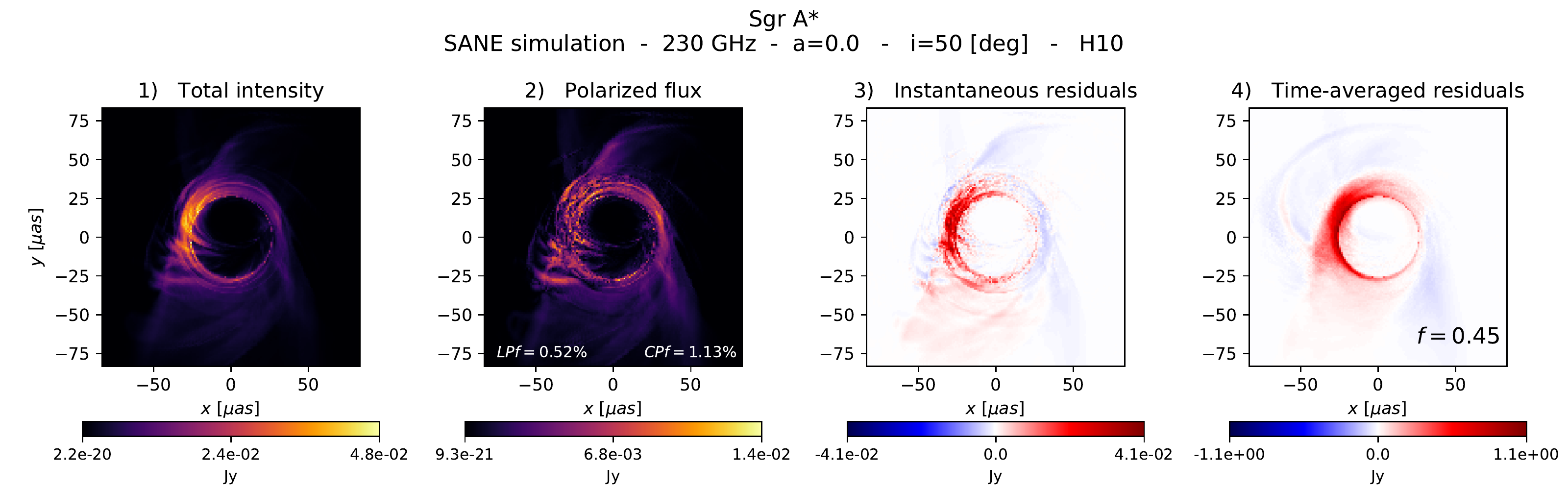}
 \caption{Total intensity, polarized flux and residual images for a Sgr A* model at 230GHz. The images are calculated from a MAD simulation with electron heating from magnetic reconnection (top) and SANE simulation with turbulent heating (bottom). In both cases, the photon ring appears as a less polarized feature in Panel 2. In the case of the MAD W18 images, the extraction method appears to work well, producing a ring with sharper features. This is unclear for the SANE case.
}
\label{fig:other_models}
\end{figure*}

Total intensity, polarized flux and residual images are shown in Fig.~\ref{fig:other_models} for both models. In both cases, the photon ring appears as a less polarized feature in the polarized flux images. In the case of the MAD W18 images, the extraction method appears to work well, with a sharp ring feature in the time-averaged residuals. This is not the case for the SANE H10, where it is unclear if photon ring features are sharper than those from total intensity images.

\subsection{Varying inclinations}

We explore different inclinations in Sgr A* images at 230 GHz using our fiducial MAD and SANE GRMHD simulations. 

\begin{figure*}
\centering
\includegraphics[trim = 4cm 9.52cm 4cm 0cm, clip=true,width=1\textwidth]{./figures_final/sgra_sane_gmin3_all_effects_230i15a0_panels.pdf}
\includegraphics[trim = 0cm 0cm 0cm 0.65cm, clip=true,width=1\textwidth]{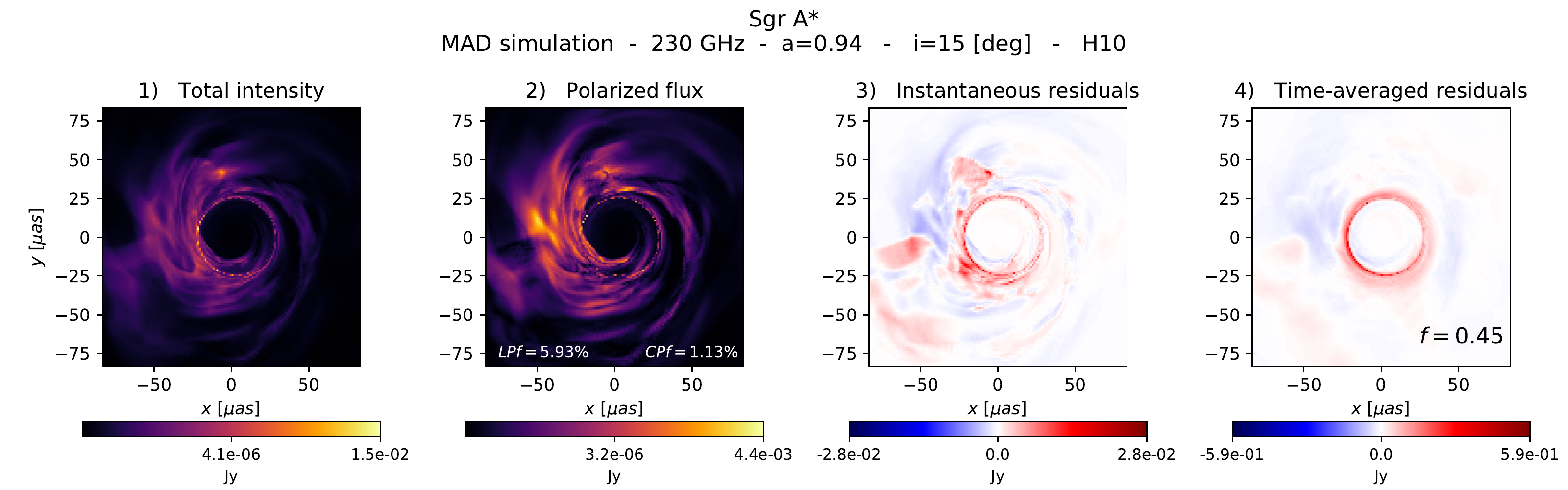}
\includegraphics[trim = 0cm 0cm 0cm 0.65cm, clip=true,width=1\textwidth]{./figures_final/sgra_mad_gmin1_all_effects_230i50a94_panels.pdf}
\includegraphics[trim = 0cm 0cm 0cm 0.65cm, clip=true,width=1\textwidth]{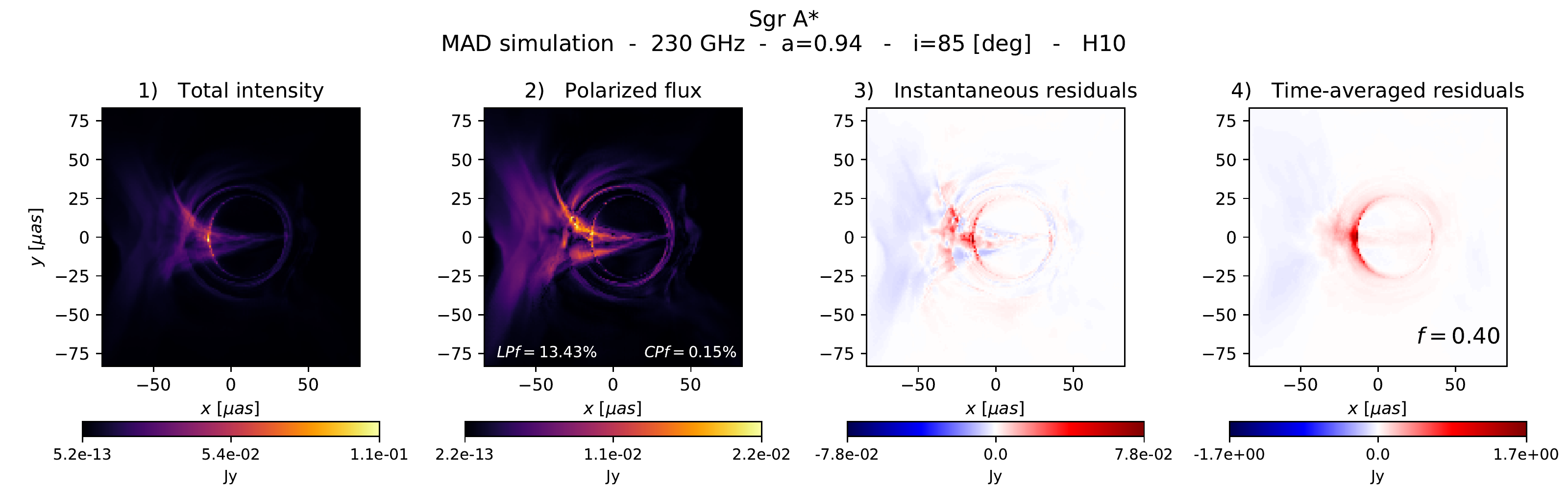}
 \caption{Comparison of Sgr A* images at 230GHz from the fiducial MAD H10 simulation at three different inclinations. Top: $i=15^\circ$. Middle:  $i=50^\circ$. Bottom: $i=85^\circ$. The method works better at lower inclinations.}
\label{fig:other_inclinations_MAD}
\end{figure*}

\begin{figure*}
\centering
\includegraphics[trim = 4cm 9.52cm 4cm 0cm, clip=true,width=1\textwidth]{./figures_final/sgra_sane_gmin3_all_effects_230i15a0_panels.pdf}
\includegraphics[trim = 0cm 0cm 0cm 0.65cm, clip=true,width=1\textwidth]{./figures_final/sgra_sane_gmin3_all_effects_230i15a0_panels.pdf}
\includegraphics[trim = 0cm 0cm 0cm 0.65cm, clip=true,width=1\textwidth]{./figures_final/sgra_sane_gmin3_all_effects_230i50a0_panels.pdf}
\includegraphics[trim = 0cm 0cm 0cm 0.65cm, clip=true,width=1\textwidth]{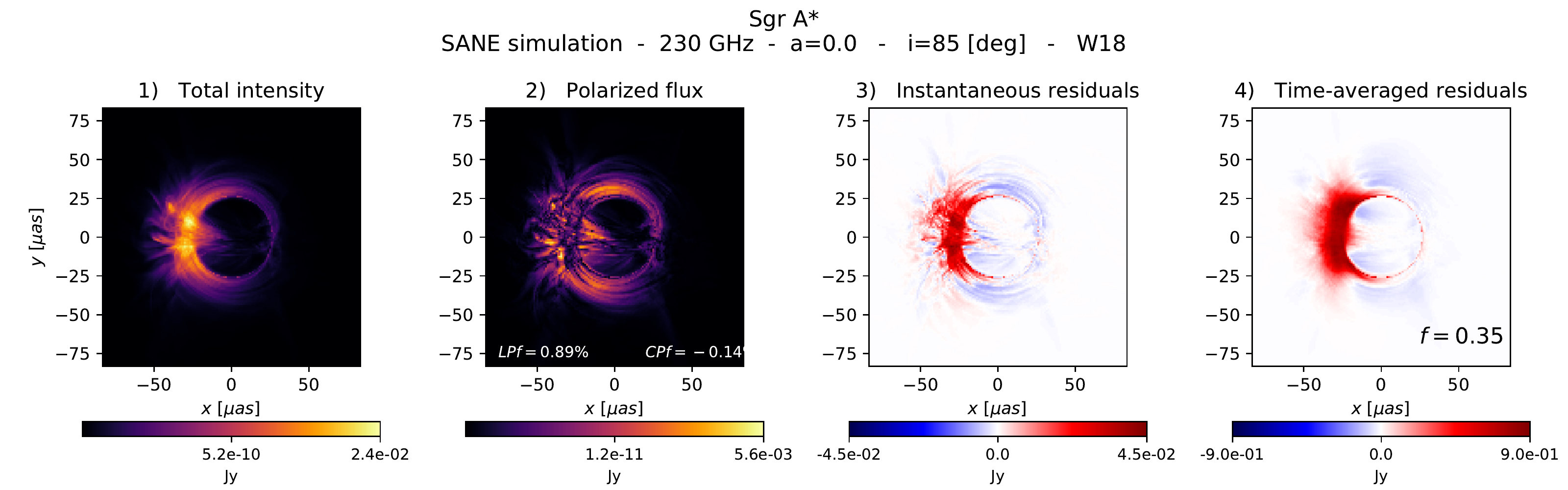}
 \caption{Comparison of Sgr A* images at 230GHz from the fiducial SANE W18 simulation at three different inclinations. Top: $i=15^\circ$. Middle:  $i=50^\circ$. Bottom: $i=85^\circ$. The method works better at lower inclinations.}
\label{fig:other_inclinations_SANE}
\end{figure*}

Total intensity, polarized flux and residual images are shown in Figures~\ref{fig:other_inclinations_MAD} and \ref{fig:other_inclinations_SANE} for the MAD and SANE models respectively. The photon ring is significantly less polarized. The method works better at lower/moderate inclinations and fails at high inclinations.

\subsection{MAD with spin a=0.0}
\label{appendix:MADa0}
We study Sgr A* images calculated from a slightly different GRMHD MAD simulation. 
Both the images parameters and GRMHD initial conditions remain the same as those described in Section~\ref{sec:method} with the only difference that the spin value is $a=0.0$. 
Just like in the fiducial MAD simulation, turbulent electron heating (H10) is considered. 

\begin{figure*}
\centering
\includegraphics[trim = 3cm 8.8cm 3cm 0cm, clip=true,width=1\textwidth]{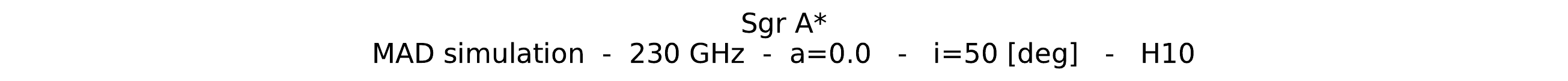}
\includegraphics[trim = 0cm 0cm 0cm 1.5cm, clip=true,width=1\textwidth]{./figures_final/sgra_mad_gmin1_all_effects_230i50a0_panels.pdf}
\includegraphics[trim = 0cm 0cm 0cm 1.55cm, clip=true, width=0.45\textwidth]{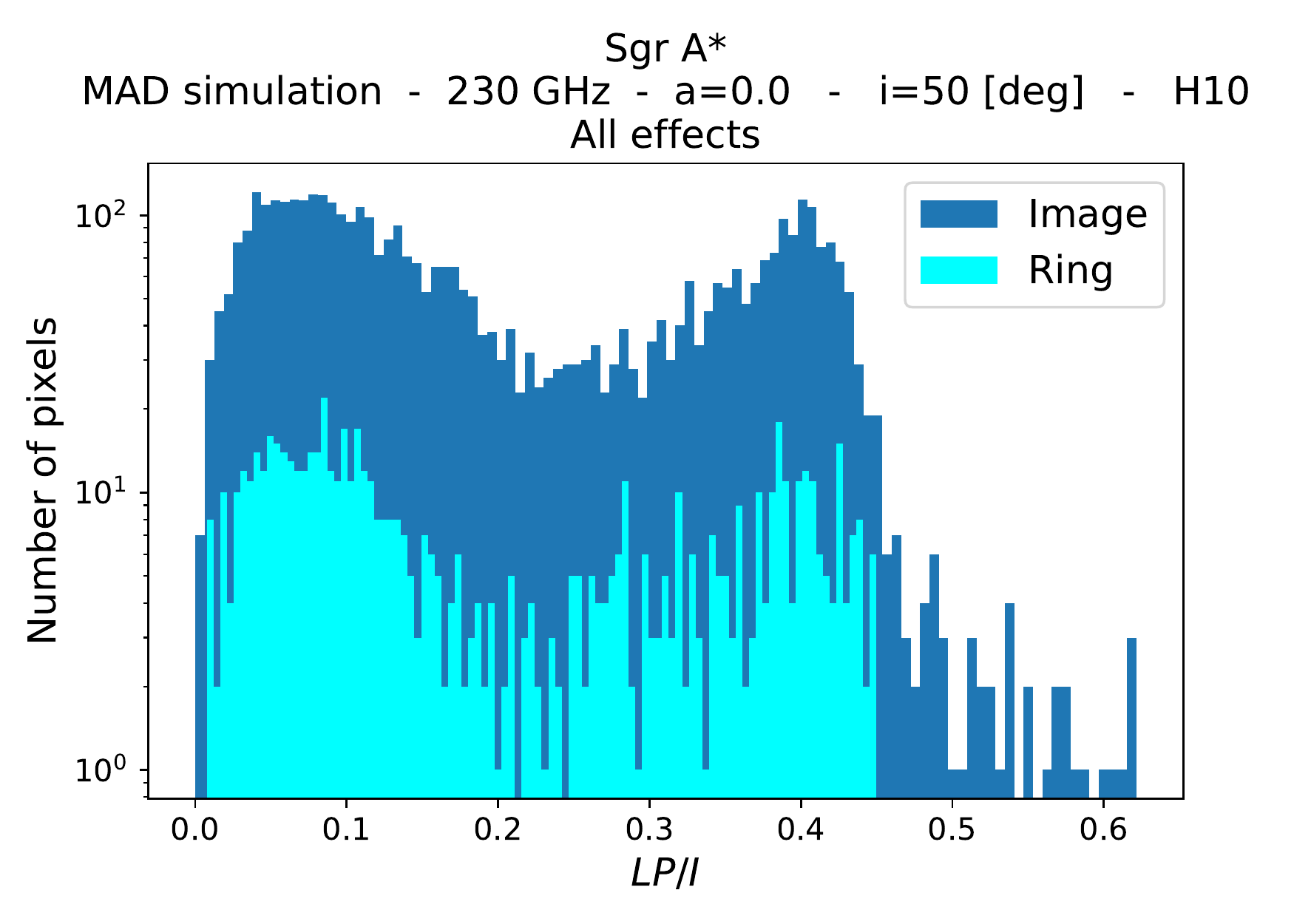}
 \caption{Sgr A* images from a MAD simulation with black hole spin $a=0.0$. Top: Total intensity, polarized flux, instantaneous residuals and time-averaged residuals. It is unclear that the photon ring is more visible in the time-averaged residual image compared to the total intensity panel. 
 Bottom: distribution of image and ring fractional polarized flux ($LP/I$) per pixel. Only pixels with a total intensity of at least $10\%\ I_{\rm{med}}$ are considered, with $I_{\rm{med}}$ the median of the one hundred brightest pixels. Image pixels are in dark blue, ring pixels are in cyan. $LP/I$ is calculated from thirty-frame averaged images where all plasma effects are considered. 
 }
     \label{fig:histograms_MADa0}
\end{figure*}

The top panel of Figure ~\ref{fig:histograms_MADa0} presents the photon ring extraction technique applied to these images. In this case it is unclear that the method produces a sharper ring than that obtained directly from total intensity images.

The bottom panel of Figure~\ref{fig:histograms_MADa0} shows the distribution of the Sgr A* MAD $a=0.0$ images with all plasma effects included. Both pixel distributions show similar behaviours.
Table~\ref{tab:fluxes_MADa0_med} shows the net LP in the image and ring pixels with a total intensity of at least $10\%\ I_{\rm{med}}$, with $I_{\rm{med}}$ the median of the one hundred brightest pixels.
The photon ring continues to be more depolarized than the rest of the image. However, $m_{\rm{Ring}}/ m_{\rm{Image}}\simeq 0.8-0.9$. Plasma effects, particularly absorption, are an important source of depolarization in this case. 

\begin{table}
  \centering
	\caption{Net fractional LP ($m$) for a Sgr A* model using a MAD simulation with spin $a=0.0$ and turbulent electron heating (H10). The values are calculated from a thirty-frame averaged image including effects of absorption and/or Faraday rotation and conversion at a time. 
	}
  \label{tab:fluxes_MADa0_med}
  \resizebox{\columnwidth}{!}{%
  \begin{tabular}{*{6}{|c}|}
  \hline
  & & All effects & No effects & No absorption & No Faraday \\
  \hline\hline
  MAD  & $m_{\rm{Image}}\ [\%]$ & 42.25 & 48.69 & 46.50 & 44.51 	\\
  a=0.0	  & $m_{\rm{Ring}}\ [\%]$  & 35.19 & 44.52 & 43.56 & 36.63  \\
     & $m_{\rm{Ring}}\ /\ m\ _{\rm{Image}} $ & 0.83  & 0.91  & 0.94  & 0.82 \\
  \hline\hline
\end{tabular}%
}
\end{table}


\bsp	
\label{lastpage}
\end{document}